\documentclass[aps,pra,twocolumn,groupedaddress,floatfix,showkeys]{revtex4-1}
\usepackage{multirow}
\usepackage{graphicx}
\usepackage{enumerate}
\usepackage{amssymb}
\usepackage{fdsymbol}
\usepackage{mathtools}
\usepackage{wasysym}
\usepackage{xcolor}

\begin{document}

\def\crta{\vrule height1.41ex depth-1.27ex width0.34em}
\def\dj{d\kern-0.36em\crta}
\def\Crta{\vrule height1ex depth-0.86ex width0.4em}
\def\Dj{D\kern-0.73em\Crta\kern0.33em}
\dimen0=\hsize \dimen1=\hsize \advance\dimen1 by 40pt

\title{Automated Generation of Arbitrarily Many {K}ochen-{S}pecker
  and Other Contextual Sets in Odd Dimensional Hilbert
  Spaces}

\author{Mladen Pavi\v ci\'c}
\email{mpavicic@irb.hr}
\affiliation{Center of Excellence for Advanced Materials 
and Sensing Devices (CEMS), Photonics and Quantum Optics Unit, 
Ru{\dj}er Bo\v skovi\'c Institute 
and Institute of Physics, Zagreb, Croatia.}

\author{Norman D. Megill}
\email{Passed away on December 9, 2021}
\affiliation{Boston Information Group, Lexington, MA 02420,
  U.S.A.}

\begin{abstract}
Development of quantum computation and communication, recently
shown to be supported by contextuality, arguably asks for a
requisite supply of contextual sets. While that has been achieved in
even dimensional spaces, in odd dimensional spaces only a dozen 
contextual critical Kochen-Specker (KS) sets have been found so far.
In this paper we give three methods for automated generation of
arbitrarily many contextual KS and non-KS sets in any dimension for
possible future application and implementation and we employ them to
obtain millions of KS and other contextual sets in dimensions 3, 5,
7, and 9 where previously only a handful of sets have been found.
Also, no explicit vectors for the original Kochen-Specker set were
known so far, while we now generate them from 24 vector components.
\end{abstract}

 \maketitle

Contextuality is paving the road for applications in quantum
computation \cite{magic-14,bartlett-nature-14}, quantum steering
\cite{tavakoli-20}, and quantum communication \cite{saha-hor-19}, as
proved by the methods of processing, preparing, and measuring of
{\em qudits\/} (quantum dits). Qudits are the units of quantum
information carried by a quantum system whose number of states
($d$) is an integer greater than one. {\em Qubits\/} (quantum
bits) are two-dimensional qudits ($d=2$) and their tensor
products build the corresponding even-dimensional Hilbert spaces.
The smallest contextual sets from these spaces have been
implemented in a series of experiments, while in 
\cite{waeg-aravind-jpa-15,pavicic-pra-17,pm-entropy18,pavicic-entropy-19,pwma-19}
billions of contextual sets in 4-, 6-, 8-, $16{\textrm -},$ and
32-dim Hilbert spaces,  predominantly related to qubits, were
generated by means of vector component algorithms (called method
{\bf M3} below), particular symmetries, geometries, polytope
correlations, parity filtering, Pauli operators, qubit states,
and dimensional upscaling.

In contrast, far fewer contextual sets based on qudits in odd-dim
spaces have been obtained so far. In particular, of Kochen-Specker
contextual sets only four in the 3-dim space
\cite{pmmm05a-corr,pavicic-entropy-19}, four in the 5-dim space
\cite{cabell-est-05},\cite[Supp.~Material]{waeg-aravind-pra-17},
five in the 7-dim space
\cite{zimba-penrose,cabell-est-05},\cite[Supp.~Material]{waeg-aravind-pra-17},
two in the 9-dim space \cite[Supp.~Material]{waeg-aravind-pra-17}, and
two in the 11-dim space \cite[Supp.~Material]{waeg-aravind-pra-17}.
General methods for automated generation of sets of a chosen
structure did not exist.

Since the quantum communication and computation are supported by
contextuality \cite{magic-14,bartlett-nature-14}, the actual
potential use of a large supply of contextual sets is twofold.
First, quantum computation algorithms which would rely on
contextual sets would arguably rely on a variety of such
sets and on their automated generation. Second, structural
properties of contextual sets differ according to their
coordinatization, parities, dimensions, sizes, etc., and
that can lead us to better understanding and applications of
the sets. 

In this paper, we offer universal and general algorithms for
automated generation of arbitrarily many contextual sets in any
dimension. In contrast to them, the programs we wrote to implement
them are computationally demanding and therefore, here, we use them
to generate sets that have not been generated so far: billions of KS
and contextual non-KS sets in $n=3,5,7,9$-dim spaces. The programs
themselves are freely available from our repository and technical
details of their previous versions are given in
\cite{waeg-aravind-jpa-15,pavicic-pra-17,pm-entropy18,pavicic-entropy-19,pwma-19}.

To describe and handle contextual sets we make use
of McKay-Megill-Pavi\v ci\'c-hypergraph (MMPH) language
\cite{pmmm05a-corr,pavicic-arxiv-unpubl-21}. An MMPH is a connected
$n$-dim {\em hypergraph} $k$-$l$ with $k$ vertices and $l$ hyperedges
in which (i) every vertex belongs to at least one hyperedge;
(ii) every hyperedge contains at least $2$ and at most $n$
vertices; (iii) no hyperedge shares only one vertex with another
hyperedge; (iv) hyperedges may intersect each other in at most
$n-2$ vertices. Graphically, vertices are represented as dots and
hyperedges as (curved) lines passing through them.

We encode MMPHs by means of the following 90 ASCII characters:
{\tt 1 2 \dots\ 9 A B \dots\ Z a b
 \dots\ z !\ " \# {\$} \% \& ' ( ) * - / : ; \textless\ =
\textgreater\ ? @ [ {$\backslash$} ] \^{} \_ {`} {\{} {\textbar} \}
$\sim$}\quad\cite{pmmm05a-corr}. When all 90 characters are
exhausted, we reuse them prefixed by `{\tt +}' (91st character),
then again by `{\tt ++}', and so on. So encoded single ASCII
characters (possibly prefixed by {\tt +}'s) represent vertices; e.g.,
{\tt 1} or {\tt +++A}. Put together they represent
hyperedges; e.g., {\tt 123} or {\tt 1+1+++1Dd}. To represent an MMPH,
hyperedges are organized in a string in which they are separated by
commas; each string ends with a period; e.g., the string
{\tt 123,345,567,789,9A1.}~represents a noncontextual 3-dim MMPH
pentagon. There is no limit on the size of an MMPH.
An MMPH is a special kind of a general hypergraph in the sense that
none of the aforementioned points (i-iv) holds for it.

Of course, instead of ASCII characters we could have used Unicode
characters or 16-bit integers but 20 years ago we decided to
proceed with the ASCII characters to encode MMPH strings and design
our algorithms and programs which in turn yield all properties and
features of MMPHs as well as their figures within the MMPH
language. All our papers since 2000 \cite{bdm-ndm-mp-1} make use
of the ASCII characters for the purpose. 

The MMPHs above are defined without a coordinatization. The
meaning of coordinatization is that a vector is assigned to
each vertex and that all vectors assigned to vertices belonging
to a common hyperedge are orthogonal to each other. Thus, 
in the MMPHs above, neither their vertices nor their hyperedges
are related to either vectors or operators. We say that an MMPH
is in an $n$-dim space, called MMPH space, when either all its
hyperedges contain $n$ vertices or when we might add vertices to
hyperedges so that each contains $n$ vertices. Orthogonality
between vertices in an MMPH space just means that they are
contained in common hyperedges. Our programs may handle MMPHs
without any reference to either vectors or projectors. In an
MMPH with a coordinatization, i.e.~with vectors assigned to
vertices, an $n$-dim MMPH space becomes an $n$-dim Hilbert
space spanned by a maximal number of (possibly added) vectors
within hyperedges. Whether we speak about an MMPH without or
with a coordinatization will be clear from the context.  

A {\em non-binary\/} MMPH (NBMMPH) is an $n$-dim
$(n\ge 3)$ $k$-$l$ MMPH, whose $i$-th hyperedge contains $\kappa(i)$
vertices $(2\le\kappa(i)\le n$,\ $i=1,\dots,l)$, to which it is
impossible to assign {\rm 1}s and {\rm 0}s in such a way that the
following rules hold: (I) no two vertices in any hyperedge are both
assigned the value $1$ and (II) in any hyperedge, not all of the
vertices are assigned the value $0$. A {\em binary\/} MMPH (BMMPH)
is an MMPH for which such an assignment is possible. NBMMPHs are
nonclassical and contextual since they do not allow assignments of
predefined 0s and 1s to their vertices. BMMPHs are classical and
noncontextual since they do allow such an assignment.
A KS MMPH is an NBMMPH with $\kappa(i)=n,\ \forall i$.
The assignments of 0s and 1s do not require a coordinatization but
an implementation of (N)BMMPHs does require it as well as their
filled MMPHs, i.e. those in which to all hyperedges with
$\kappa(i)\le n$ we add $n-\kappa(i)$ vertices so as to satisfy the
mutual orthogonalities. An example of a non-KS NBMMPH without a
coordinatization is the 33-27 in Fig.~1(d) in the Supplemental
Material (SM).

When either state-dependent or
state-independent tests of operators defined on vertices of an
NBMMPH with $\kappa(i)\le n$ confirm the contextuality,
e.g.~\cite{yu-oh-12,cabello-bengtsson-12,cabello-svozil-18},
then the NBMMPH turns out to be contextual in all considered cases
so far. 

A {\em critical\/} NBMMPH is an NBMMPH which after removing any of
its hyperedges becomes a BMMPH. 

To generate (N)BMMPHs in the odd-dim spaces we make use of
three methods---{\bf M1-M3}.

{\bf M1} consists in an automated dropping of vertices contained in
single hyperedges (multiplicity $m=1$) \cite{pavicic-arxiv-unpubl-21}
of MMPHs and a possible subsequent stripping of their hyperedges.
The obtained smaller MMPHs are often NBMMPH although never KS.  

{\bf M2} consists in an automated random addition of hyperedges to
MMPHs so as to obtain larger ones which then serve us to generate
smaller KS MMPHs by stripping hyperedges randomly again;

{\bf M3} consists in combining simple vector components so as to
exhaust all possible collections of $n$ mutually orthogonal $n$-dim
vectors. These form big {\em master\/} MMPHs which consist of single
or multiple MMPHs of different sizes. Master MMPHs may or may not be
NBMMPH, what we find out by applying filters to them.
NBMMPHs serve us to massively generate a {\em class} of
smaller MMPHs via our algorithms and programs.

We carry out methods {\bf M1-M3} and by means of our programs
{\textsc{MMPSstrip}} (for stripping and adding hyperedges),
{\textsc{States01}} (for verifying the contextuality),
{\textsc{MMPShuffle}} (for reorganizing MMPHs), and 
{\textsc{ONE}} (for evaluating the structural properties of
MMPHs \cite{pavicic-arxiv-unpubl-21}).
  
We combine all three methods to obtain a large number of
NBMMPHs in odd-dim spaces.

\begin{figure}[ht]
\center
\includegraphics[width=0.48\textwidth]{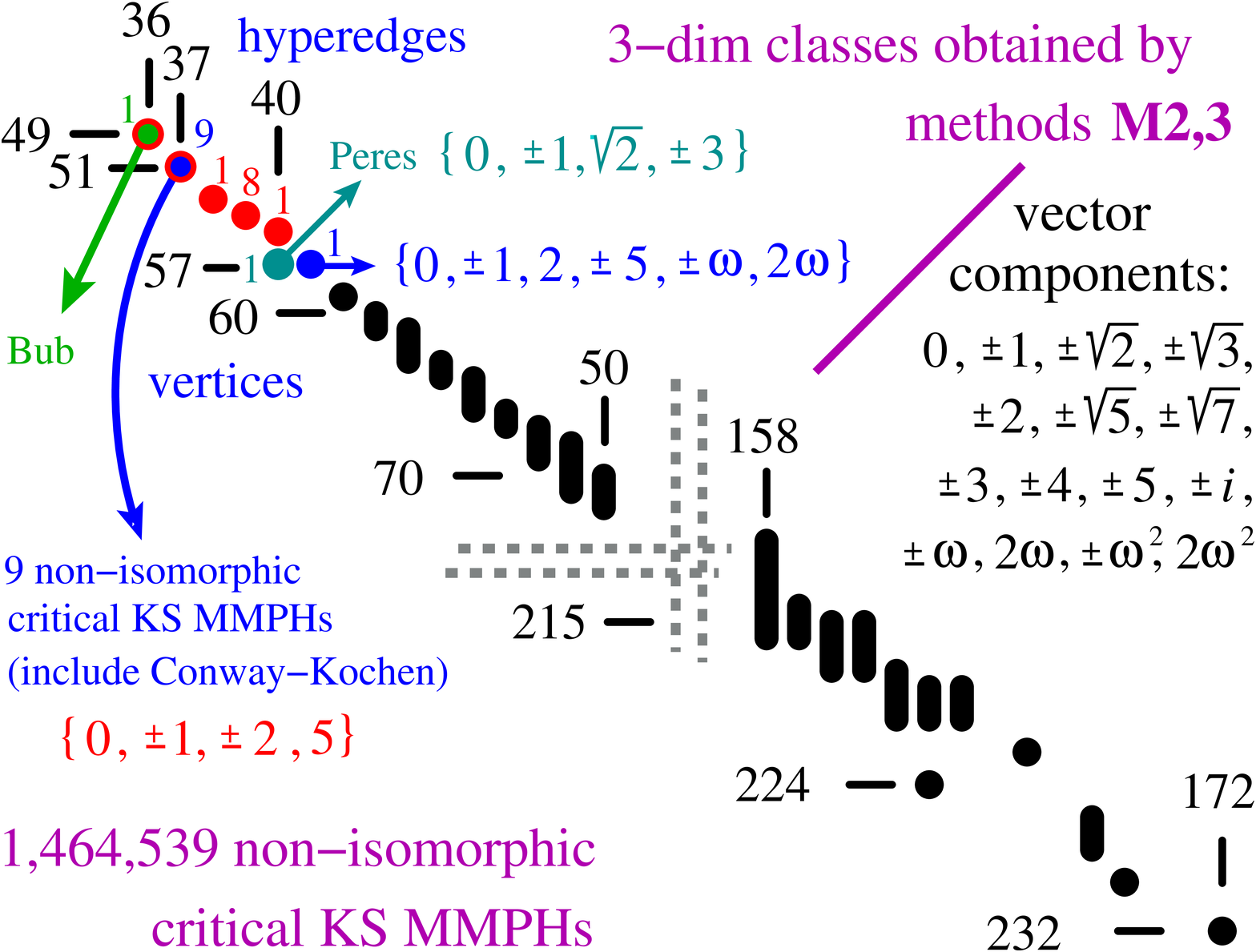}
\caption{The distribution of critical 3-dim KS MMPHs obtained
  via {\bf M2,3} with the given vector components.
  Abscissa is $l$ (number of hyperedges); ordinate is $k$
  (number of vertices). Dots represent $(k,l)$. Consecutive dots
  (same $l$) are shown as strips. The same applies to
  Figs.~\ref{fig:5d} and \ref{fig:7d}. See text.
  MMPH strings, figures, and vectors are given in SM.}
\label{fig:3d}
\end{figure}

{\bf 3-dim case.} So far there have been only four known KS MMPHs,
and as we show in \cite{pavicic-entropy-19} none of their varieties
with stripped $m=1$ vertices is critical. Via {\bf M1}, i.e.~by
stripping their edges, we can obtain thousands of smaller non-KS
NBMMPHs and BMMPHs down to a pentagon\cite{pavicic-entropy-19}. But
this means that they are limited in size by the size of four
original MMPHs. To get larger MMPHs we have to apply {\bf M2,3}.
The final distribution of critical KS MMPHs we generated is given
in Fig.~\ref{fig:3d}. 

{\bf M2} consists in adding hyperedges to Bub, Conway-Kochen,
and Peres' KS MMPHs (for citations and figures see
\cite{pavicic-pra-17}) using our program {\textsc{MMPSstrip}},
filtering out KS MMPHs, and stripping them to the critical KS
MMPHs by {\textsc{States01}}. The latter critical MMPHs build
their coordinatization from the two vectors component sets of
the original three MMPHs.

These vector components also serve us to obtain the same critical
KS MMPHs by employing {\bf M3} so as to exhaust all possible
collections of 3 mutually orthogonal vectors representing
hyperedges interwoven in MMPHs. 

When applying {\bf M3} we obtain that Bub's is the only 49-36 NBMMPH
and that there are no smaller KS ones for the considered vector
components. There are no other critical KS MMPHs between 49-36 and
51-37; between 51-37 and Peres' 57-40 we obtained the following
non-isomorphic MMPHs: one 53-38, eight 54-39, and one 55-40. Peres'
57-40 is generated from the components $\{0,\pm1,\pm\sqrt{2},3\}$
and all the smaller ones from $\{0,\pm1,\pm2,5\}$.
When we apply {\bf M2} so as to add sufficiently many
hyperedges to any of the three original MMPHs (Bub, Conway-Kochen,
or Peres') and then strip them back down in a search for smaller
critical MMPHs, we always obtain the other two among the generated
critical KS MMPHs.

The more components we use with {\bf M3}, the larger the master
files and the more critical KS MMPHs we obtain. For instance
with the help of $\{0,\pm 1,\sqrt{2},3\}$ (Peres') components 
we obtain the master 81-52 which contains just one single
critical set---Peres' 57-40; $\{0,\pm 1,\pm 2,5\}$ yield the
master 97-64 which generates 20 critical KS MMPHs from 49-36 to
55-40; in contrast, $\{0,\pm 1,\sqrt{2},\pm 2,\pm 3,5\}$ yield the
master 301-184 which generates 81 critical KS MMPHs from 49-36 to
92-66; $\{0,\pm\omega,2\omega,\pm\omega^2,2\omega^2\}$, where
$\omega$=$e^{2\pi i/3}$, yield 514 criticals from 69-50 to 106-79,
etc. Several smallest MMPHs from these classes are shown in
Table \ref{T:small} and SM.

\begin{table}[ht]
\center
\setlength{\tabcolsep}{0.5pt}
\begin{tabular}{|c|ccccc|}
    \hline
  \multirow{3}{*}{\rotatebox{90}{dimen.\ }} & Smallest &
  No.~of  & \multirow{3}{*}{\rotatebox{90}{Methods}}
  & \multirow{2}*{Smallest}
  & \multirow{2}*{Vector} \\
&  critical   &  non-isom & & \multirow{2}*{master} &
      \multirow{2}*{components} \\
&MMPHs        & MMPHs    &  &     & \\
  \hline
  \multirow{8}{*}{\rotatebox{90}{3D MMPHs\ }}& 5-5$^\dag$ & 1 &
 {\bf M1}& 40-23 & $\{0,\pm 1,2\}$ \\
  &  19-13 & 1 & {\bf M2}& 20-14 & none \\
  &  39-27 & 1 & {\bf M2}& 39-30 & ? \\
  &  49-36 & 1 & \multirow{3}*{$\!\!\!\!\!\!\!\begin{rcases*} \\ \\ \\ \end{rcases*}$\bf \ M2-3} & & \\
 &  51-37 & 9 &  & 97-64 & $\{0,\pm 1,\pm 2,5\}$\\
 &    53-38 & 1 &  &  & \\
&57-40 & 1 & {\bf M2-3}& 81-52 & $\{0,\pm 1,\sqrt{2},\pm 3\}$ \\
&  69-50 & 3 & {\bf M3}& 169-120 &
  $\!\!(0,\pm\omega,2\omega,\pm\omega^2,2\omega^2\}$ \\
  \hline
  {\rotatebox{90}{$\!\!$5D}}&29-16& 2 & {\bf M3}& 105-136 & 
   $\{0,\pm 1\}$ \\  
  \hline
\multirow{4}{*}{\rotatebox{90}{7D\quad}}
  &13-4& 1 &{\bf M1,2}&24-6& ?\\
  &28-14$^\dag$& 1 &{\bf M2,3}&805-9936& $\{0,\pm 1\}$\\
  &34-14& 1 &{\bf M3}&805-9936& $\{0,\pm 1\}$\\
    &207-97& 1 & {\bf M3}& 805-9936 & $\{0,\pm 1\}$ \\
   \hline
  \multirow{3}{*}{\rotatebox{90}{9D\ \ }}&19-5& 2 & {\bf M2}& 20-6 & ? \\
  &47-16& 2 & {\bf M3}& 9586-12068704 & $\{0,\pm 1\}$ \\
  &7-6$^\dag$& 1 & {\bf M1,3}& 9586-12068704 & $\{0,\pm 1\}$ \\  
  \hline\end{tabular}
\caption{Some of smaller NBMMPHs obtained by methods {\bf M1-3};
  `$\dag$' indicates that the MMPH is a non-KS NBMMPH---all the
  others are KS NBMMPHs; `?' indicates that obtaining the
  coordinatization is too demanding and that we were not able to
  carry it out on our supercomputers or that it might not exist
  at all.}
\label{T:small}
\end{table}

We did not find simple real vector components which would yield
KS MMPHs smaller than 49-36, although we are able to generate many
smaller KS MMPHs down to 19-13 or 39-27 shown in SM, although
without a coordinatization based on components shown in
Fig.~\ref{fig:3d}.

The path taken in \cite{vor-nav-21} is intractable for hundreds of
small KS MMPHs we checked on our supercomputer since the number
of free variables is too high. See Fig.~1(a) in SM.

Another path was taken in 2021 by Jean-Pierre Merlet who applied
the interval analysis method of solving nonlinear equations to the
19-13 MMPH. The equations turned out not to have a real solution and
complex ones were not calculable even on a supercomputer.

As for possible coordinatizations of smaller KS MMPHs, we draw a
parallel with the original KS set 192-118
\cite{koch-speck}. Its trigonometric formula in \cite{koch-speck}
looks simple, but its coordinatization is so complicated that a
random search for them is infeasible. More specifically, the
aforementioned trigonometric formula is not sufficient to provide
a coordinatization. We, therefore, constructed apparently the first
known coordinatization of the original Kochen-Specker set  
from 24 components in SM.

\begin{figure}[ht]
\center
\includegraphics[width=0.48\textwidth]{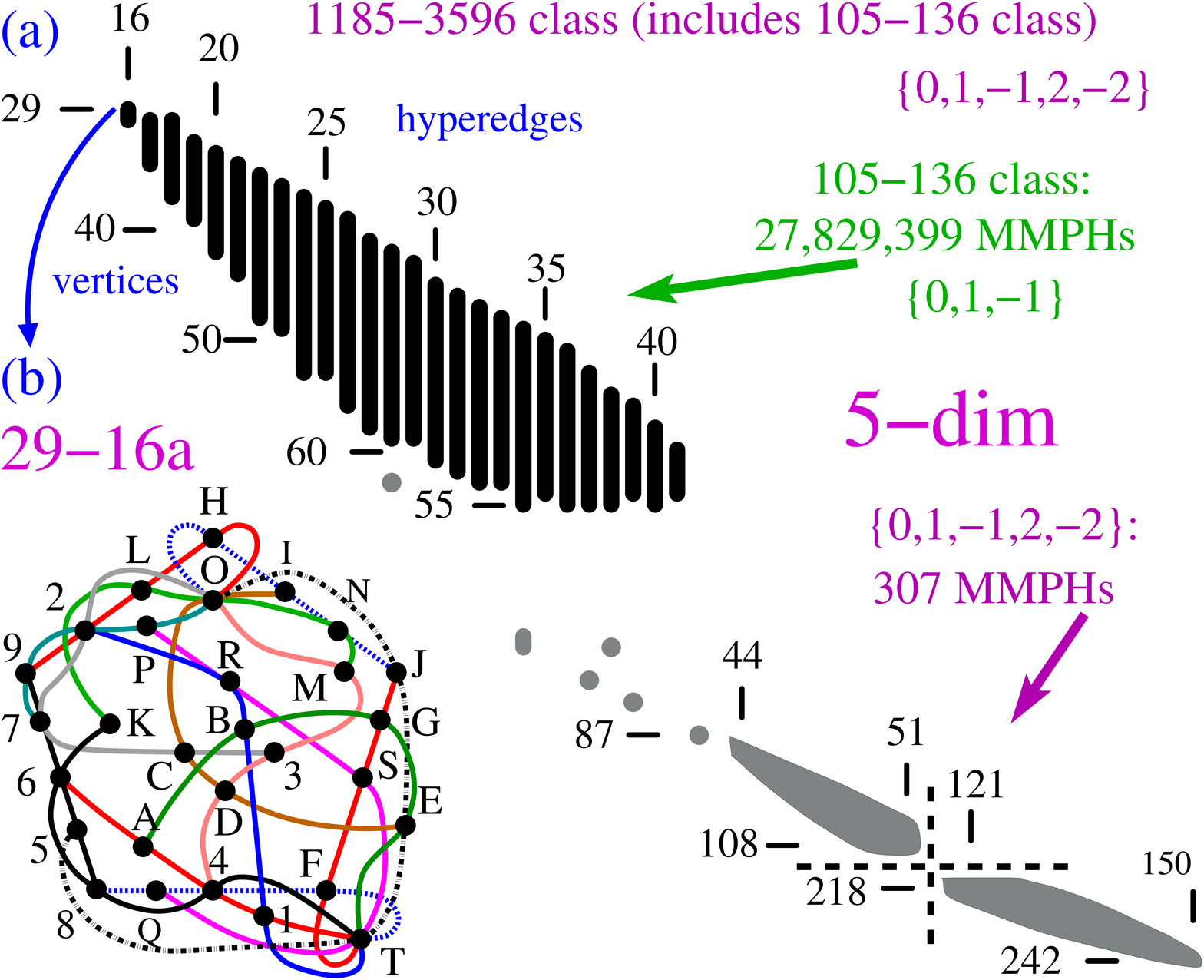}
\caption{(a) Distribution of the 5-dim critical KS MMPHs obtained
  by {\bf M3}; the generation of the 1185-3596 was demanding and
  CPU time consuming, so, we stopped it after obtaining the upper
  samples which enable us to estimate the size of the class;
  (b) One of the two smallest 5-dim 29-16 critical KS MMPH; the
  other (29-16b) was previously obtained by the dimensional
  upscaling method in \cite{waeg-aravind-pra-17}.}
\label{fig:5d}
\end{figure}

{\bf 5-dim case.} In contrast to the 3-dim case, the 5-dim
KS MMPHs can be obtained from just three vector components
$\{0,\pm1\}$, which by method {\bf M3} generate the 105-136
master set and its 105-136 class of KS MMPHs. These include
critical ones from the smallest 29-16 to the biggest 64-41,
altogether 27,829,399 non-isomorphic MMPHs. The distribution
of the MMPHs within the class is shown in Fig.~\ref{fig:5d}(a)
and the 29-16 in Fig.~\ref{fig:5d}(b).

\begin{figure}[ht]
  \center
\includegraphics[width=0.48\textwidth]{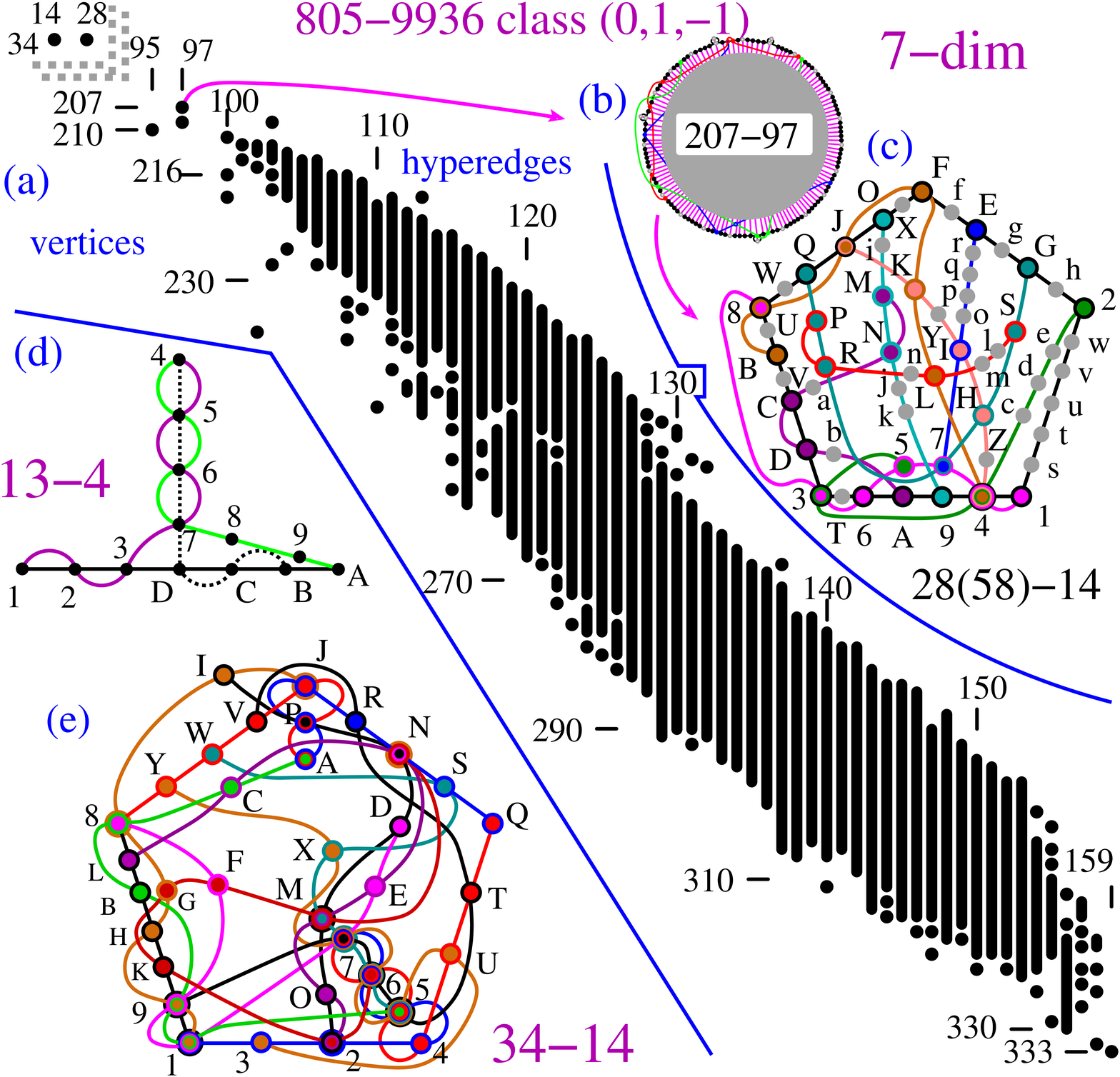}
\caption{(a) Distribution of 42,816 7-dim critical KS MMPHs from
  the class 805-9936 generated by {\bf M2,3}; the 
  master is obtained from $\{0,\pm 1\}$; (b) 207-97; too interwoven
  to be graphically represented in detail; (c) the non-KS NBMMPH
  obtained from 207-97 via {\sc MMPShuffle}; (d) critical KS 13-4
  obtained by {\bf M1,2}; (e) critical KS 34-14 MMPH obtained by
  massive targeting MMPHs with 34 vertices---see text.}
\label{fig:7d}
\end{figure}

We also obtained the master (1185-3596) and a number of elements
from its class from the five components $\{0,\pm1,\pm2\}$ but we
stopped the generation of KS criticals after only 307 MMPHs.
Their distribution is indicated in Fig.~\ref{fig:5d}---if
fully generated it would completely include the 105-136 class
and would be continuously spread over the whole ranged of
hyperedges and vertices.

{\bf 7-dim case.} We generated the 7-dim critical KS MMPHs from
the vector components $\{0,\pm1\}$ via {\bf M3} so as to first
obtain the 805-9936 master which in turns generated the the 805-9936
class shown in Fig.~\ref{fig:7d}(a). Their vectors were automatically
generated from the master set by means of our programs
{\textsc{MMPStrip, MMPShuffle}}, and {\textsc{States01}}.

The generation provided MMPHs from 207-97 to 333-159 after running
200 parallel jobs on a supercomputer for two weeks.
Longer runs would give us smaller MMPHs as proved by the 34-28 MMPH
(see the top of Fig.~\ref{fig:7d}(a)) obtained in \cite{cabell-est-05}. 
Particular targeted runs might give us particular smaller KS MMPHs,
e.g.~the critical 34-14 (Fig.~\ref{fig:7d}(a,e) and SM); it
required stripping the master down to the MMPHs with 34 vertices
via \textsc{MMPStrip} and then filtering them for the KS feature
via {\textsc{States01}; since such a procedure is too CPU-time
demanding, a search for further smaller MMPHs is out of the scope
of this paper. Instead, {\bf M1} can serve for a massive automated
generation of smaller non-KS NBMMPHs with the help of
\textsc{MMPShuffle} and \textsc{MMPStrip}. An outcome (28-14) is
shown in Fig.~\ref{fig:7d}(c), as obtained from the 207-97
Fig.~\ref{fig:7d}(a,b). The procedure is analogous to stripping
original 3-dim MMPHs \cite{pavicic-entropy-19}.

Yet another way of automated generation of 7-dim MMPHs is via
{\bf M2} by means of \textsc{MMPStrip}, an example of which is
13-6 shown in Fig.~\ref{fig:7d}(d); we were not able to find its
coordinatization and we conjecture that it does not have any.
We confirmed that it is not determined by the vector components
$\{0,\pm 1\}$ and we work on a program which would calculate
coordinatization for bigger instances of such MMPHs we generated. 

{\bf 9-dim case.} Two entangled qutrits live in a 9-dim space
and we generated the MMPH master from $\{0,\pm1\}$ components.
It consists of 9,586 vertices and 12,068,705 hyperedges and that
proved to be too huge for a direct generation of critical MMPHs
(via stripping and filtering) from the master MMPH although the
KS 47-16 given in SM proves that the master is a KS MMPH. 

\begin{figure}[ht]
\center
\includegraphics[width=0.48\textwidth]{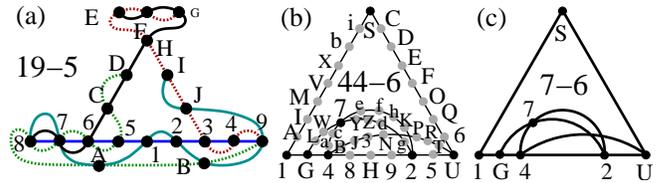}
\caption{(a) KS MMPH 19-5; no coordinatization was found; 
(b) BMMPH 44-6; (c) NBMMPH 7-6; MMPH strings (a-c) and 
vectors (b,c) are given in SM.}
\label{fig:9d}
\end{figure}

However, smaller critical KS MMPH can be obtained from simple
BMMPHs via {\bf M2}, in particular via an automated procedure
of adding hyperedges and then generating critical KS MMPHs by
stripping hyperedges via {\bf M1} and filtering them for the
critical KS property. The critical KS MMPH 19-5 obtained in this
way (Fig.~\ref{fig:9d}(a)) has no coordinatization from
$\{0,\pm 1\}$ and we conjecture that it does not have any, but it
represents the proof of principle of how {\bf M2} works.

Also, a great many of MMPHs stripped of $m=1$ vertices exhibit
contextuality. Smaller ones can easily be implemented. The smallest
one we obtained by {\bf M3,1} is shown in Fig.~2(c) of SM and
referred to in Table \ref{T:small}. Its filled MMPH is shown in
Fig.~2(c) of SM. Their differences are discussed in SM.

{\em To summarize\/}, in this paper we give methods for generating
KS as well as non-KS NBMMPHs in odd-dimensional Hilbert spaces.
Our goal is not to find ``record'' smallest MMPHs but to establish
general methods for automated generation of NBMMPHs in any dimension
for any possible future application and implementation, e.g., in
quantum computation and  communication. The methods are especially
needed in odd dimensional Hilbert spaces since, in contrast to even
dimensional ones, we cannot make use of polytopes, Pauli operators,
qubit states, parities, and other approaches specific to qubit spaces.
We propose three such methods: {\bf M1} which consists in dropping
vertices contained in single hyperedges, {\bf M2} consists in random
addition of hyperedges to MMPHs, and {\bf M3} which consists in
combining simple components so as to exhaust all possible collections
of mutually orthogonal vectors. Automated generation is achieved by
means of our algorithms and programs presented above. 

In the 3-dim space we generated roughly a million and a half
non-isomorphic KS ones, ranging from MMPHs with 19 vertices and 13
hyperedges (19-13) without a coordinatization (via {\bf M3}), over 
eleven 51-37s, up to a 232-172, all with coordinatizations, 
distributed as shown in Fig.~\ref{fig:3d}. Special cases are 
given in Table \ref{T:small} and SM. In SM we also give, for the
first time, an explicit coordinatization of the original KS set.

In the 5-dim space, from the vector components $\{0,\pm1\}$, 
we generated roughly 28 million KS MMPHs whose distribution is 
shown in Fig.~\ref{fig:5d}(a), ranging from 29-16 to 242-131.

In the 7-dim space the components $\{0,\pm1\}$ generate roughly
30,000 rather big and computationally demanding KS MMPHs shown
in Fig.~\ref{fig:7d}(a), ranging from 34-14 to 333-159.

The 9-dim MMPH master generated by $\{0,\pm1\}$ has 9,586 vertices
and 12,068,705 hyperedges and it is computationally too demanding
to yield critical KS MMPHs directly.

We also explain how a combination of methods {\bf M1-M3} can
be employed to generate targeted smaller classes of non-KS
NBMMPHs. This approach is important because the operator and
projector based contextual sets, which are recently being used
in the literature, are often built by means of such NBMMPHs. 

\begin{acknowledgments}
  Supported by the Ministry of Science and Education of Croatia
  through the Center of Excellence CEMS funding, and by MSE grants
  Nos.~KK.01.1.1.01.0001 and 533-19-15-0022. Computational support
  provided by the Zagreb University Computing Centre. Repository
{\tt http://puh.srce.hr/s/Qegixzz2BdjYwFL}
\end{acknowledgments}

\begin{widetext}

\section*{Supplemental Material}

  \subsection*{3-dim}
  
  \subsubsection*{One of eight  {\rm 51-37 MMPH}s generated by
    $\{0,\pm 1,\pm 2,5\}$, nonisomorphic to Conway-Kochen's
    {\rm MMPH}}

  {\bf 51-37} {\tt 213,35b,bYh,hgR,RQN,N6C,CBD,DaF,FEG,GHI,IOP,PTU,UpK,KJL,LV8,89A,Akl,lcf,fde,eZX,XWn,nm2,456,\break 783,MNL,STR,WDA,XYK,ZaV,ZI6,cG2,cYC,WT5,cVT,ijP,oSF,jba.} {\tt 1}=(-1,5,2); {\tt 2}=(2,0,1); {\tt 3}=(1,1,-2); {\tt 4}=(1,1,2);\break {\tt 5}=(1,-1,0); {\tt 6}=(1,1,-1); {\tt 7}=(5,-1,2); {\tt 8}=(0,2,1); {\tt 9}=(5,1,-2); {\tt A}=(1,-1,2); {\tt B}=(1,2,-1); {\tt C}=(1,0,1); {\tt D}=(-1,1,1); {\tt E}=(-2,5,-1);\break {\tt F}=(2,1,1); {\tt G}=(1,0,-2); {\tt H}=(2,5,1); {\tt I}=(2,-1,1); {\tt J}=(5,-2,-1); {\tt K}=(1,2,1); {\tt L}=(0,1,-2); {\tt M}=(5,2,1); {\tt N}=(-1,2,1); {\tt O}=(-2,1,5); {\tt P}=(1,2,0); {\tt Q}=(1,-2,5); {\tt R}=(2,1,0); {\tt S}=(1,-2,0); {\tt T}=(0,0,1); {\tt U}=(2,-1,0); {\tt V}=(1,0,0); {\tt W}=(1,1,0); {\tt X}=(1,-1,1); {\tt Y}=(1,0,-1); {\tt Z}=(0,1,1); {\tt a}=(0,1,-1); {\tt b}=(1,1,1); {\tt c}=(0,1,0); {\tt d}=(-2,5,1); {\tt e}=(2,1,-1); {\tt f}=(1,0,2); {\tt g}=(-1,2,5); {\tt h}=(1,-2,1); {\tt i}=(2,-1,5); {\tt j}=(-2,1,1); {\tt k}=(1,5,2); {\tt l}=(2,0,-1); {\tt m}=(1,5,-2); {\tt n}=(-1,1,2); {\tt o}=(-2,-1,5); {\tt p}=(-1,-2,5).

  \subsubsection*{Chosen MMPHs}

\begin{figure}[ht]
\center
\includegraphics[width=0.99\textwidth]{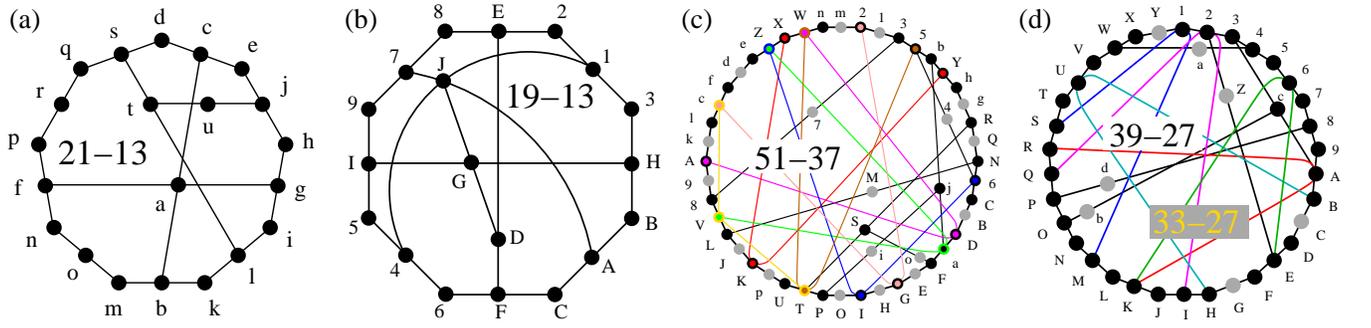}
\caption{(a) BMMPH obtained by Vor{\'a}{\v c}ek and Navara
  \cite{vor-nav-21}; the 18-13 MMPH which one obtains by
  removing the vertices {\tt o,r,u} is a NBMMPH;
  (b) KS MMPH which does not have a real coordinatization and
  most probably also not a complex one;
  (c) One of nine 51-37 KS MMPHs we generated; it is a 22-gon
  non-isomorphic to Conway-Kochen's 51-37; (d) KS NBMMPH 39-27
  does not have a known coordinatization---non-KS NBMMPH 33-27
  obtained by removal of all gray vertices from the 39-27
  therefore also does not have a known coordinatization.}
\label{fig:3d-s}
\end{figure}

\subsubsection*{39-27 critical {\rm KS MMPH}---possible
  coordinatization is an open problem}

{\bf 39-27} 1st 17 triads build a 17-gon: {\tt 837,7NF,FEG,GLK,K4c,cbY,Y2S,SRT,TWX,X6H,HJI,IQV,Vd5,5UA,A9B,BDC,CZ8,\break 123,456,MND,OPQ,UJ8,aVL,E62,dW3,P74,cW9.}

\bigskip 
  
\subsubsection*{Other {\rm MMPH}s generated by
    $\{0,\pm 1,\pm 2,5\}$}

{\bf 53-38} {\tt 213,39A,AFG,GpB,BNX,XWY,YdK,KVf,fe5,546,678,8ED,DIr,rqO,OLP,Pkl,lCH,HMa,aZb,bhJ,JSj,ji2,BC5,\break HI2,JEC,KIG,L63,MNL,LKJ,QRS,TUV,cd8,ghA,mnP,opO,nFE,UND,RMF.} {\tt 1}=(5,2,1);  {\tt 2}=(-1,2,1); {\tt 3}=(0,1,-2);\break {\tt 4}=(5,-1,2); {\tt 5}=(1,1,-2); {\tt 6}=(0,2,1); {\tt 7}=(5,1,-2); {\tt 8}=(1,-1,2); {\tt 9}=(5,-2,-1); {\tt A}=(1,2,1); {\tt B}=(1,1,1); {\tt C}=(1,-1,0); {\tt D}=(-1,1,1); {\tt E}=(1,1,0); {\tt F}=(1,-1,1); {\tt G}=(1,0,-1); {\tt H}=(1,1,-1); {\tt I}=(1,0,1); {\tt J}=(0,0,1); {\tt K}=(0,1,0); {\tt L}=(1,0,0); {\tt M}=(0,1,1); {\tt N}=(0,1,-1); {\tt O}=(0,1,2); {\tt P}=(0,2,-1); {\tt Q}=(2,1,5); {\tt R}=(2,1,-1); {\tt S}=(1,-2,0); {\tt T}=(-2,5,-1); {\tt U}=(2,1,1); {\tt V}=(1,0,-2); {\tt W}=(2,5,-1); {\tt X}=(-2,1,1); {\tt Y}=(1,0,2); {\tt Z}=(-2,1,5); {\tt a}=(2,-1,1); {\tt b}=(1,2,0); {\tt c}=(1,5,2); {\tt d}=(2,0,-1); {\tt e}=(-1,5,2); {\tt f}=(2,0,1); {\tt g}=(-1,-2,5); {\tt h}=(2,-1,0); {\tt i}=(1,-2,5); {\tt j}=(2,1,0); {\tt k}=(5,-1,-2); {\tt l}=(1,1,2); {\tt m}=(5,1,2); {\tt n}=(-1,1,2); {\tt o}=(5,2,-1); {\tt p}=(1,-2,1); {\tt q}=(5,-2,1); {\tt r}=(1,2,-1).

\medskip

One of the eight 54-39 MMPHs:

\smallskip

{\bf 54-39} {\tt 546,6DE,EmW,WRV,VUJ,JHI,Ipq,qTs,srG,GFC,CAB,B38,879,9ZL,LMN,NOP,PbY,Yci,ihj,jdg,gef,fXa,a25,\break123,KLJ,QRP,STN,XYG,bI3,cE9,cT2,dC6,dbZ,XV8,dVT,klR,nSB,oU5,laZ.} {\tt 1}=(-1,1,2); {\tt 2}=(1,1,0); {\tt 3}=(1,-1,1); {\tt 4}=(5,1,-2); {\tt 5}=(1,-1,2); {\tt 6}=(0,2,1); {\tt 7}=(1,-2,1); {\tt 8}=(1,0,-1); {\tt 9}=(1,1,1); {\tt A}=(5,-2,-1); {\tt B}=(1,2,1); {\tt C}=(0,1,-2); {\tt D}=(5,-1,2); {\tt E}=(1,1,-2); {\tt F}=(5,2,1); {\tt G}=(-1,2,1); {\tt H}=(-2,5,1); {\tt I}=(2,1,-1); {\tt J}=(1,0,2); {\tt K}=(2,5,-1); {\tt L}=(-2,1,1); {\tt M}=(2,-1,5); {\tt N}=(1,2,0); {\tt O}=(-2,1,5); {\tt P}=(2,-1,1); {\tt Q}=(2,5,1); {\tt R}=(1,0,-2); {\tt S}=(2,-1,0); {\tt T}=(0,0,1); {\tt U}=(2,0,-1); {\tt V}=(0,1,0); {\tt W}=(2,0,1); {\tt X}=(1,0,1); {\tt Y}=(1,1,-1); {\tt Z}=(0,1,-1); {\tt a}=(-1,1,1); {\tt b}=(0,1,1); {\tt c}=(1,-1,0); {\tt d}=(1,0,0); {\tt e}=(5,-2,1); {\tt f}=(1,2,-1); {\tt g}=(0,1,2); {\tt h}=(5,-1,-2); {\tt i}=(1,1,2); {\tt j}=(0,2,-1); {\tt k}=(-2,5,-1); {\tt l}=(2,1,1); {\tt m}=(-1,5,2); {\tt n}=(-1,-2,5); {\tt o}=(1,5,2); {\tt p}=(2,1,5); {\tt q}=(1,-2,0); {\tt r}=(1,-2,5); {\tt s}=(2,1,0).

\medskip
{\bf 55-40} {\tt 123,456,789,AB6,CD9,EF3,GHB,IJD,KLF,MN9,OP6,QR3,SNH,TRJ,UPL,VW5,XY8,Za2,bcB,dcD,ecF,eaN,dYP,\break bWR,RPN,fgS,hiU,jkT,lmH,noJ,pqL,rM5,sQ8,tO2,mXW,qaV,oZY,gdV,keX,ibZ.} {\tt 1}=(5,-1,2), {\tt 2}=(1,1,-2), {\tt 3}=(0,2,1), {\tt 4}=(2,5,1), {\tt 5}=(2,-1,1), {\tt 6}=(1,0,-2), {\tt 7}=(-1,-2,5), {\tt 8}=(1,2,1), {\tt 9}=(2,-1,0), {\tt A}=(-2,5,-1), {\tt B}=(2,1,1), {\tt C}=(1,2,5), {\tt D}=(1,2,-1), {\tt E}=(5,1,-2), {\tt F}=(1,-1,2), {\tt G}=(-2,-1,5), {\tt H}=(1,-2,0), {\tt I}=(5,-2,1), {\tt J}=(0,1,2), {\tt K}=(1,5,2), {\tt L}=(2,0,-1), {\tt M}=(1,2,0), {\tt N}=(0,0,1), {\tt O}=(2,0,1), {\tt P}=(0,1,0), {\tt Q}=(0,1,-2), {\tt R}=(1,0,0), {\tt S}=(2,1,0), {\tt T}=(0,2,-1), {\tt U}=(1,0,2), {\tt V}=(1,1,-1), {\tt W}=(0,1,1), {\tt X}=(1,-1,1), {\tt Y}=(1,0,-1), {\tt Z}=(1,1,1), {\tt a}=(1,-1,0), {\tt b}=(0,1,-1), {\tt c}=(-1,1,1), {\tt d}=(1,0,1), {\tt e}=(1,1,0), {\tt f}=(1,-2,5), {\tt g}=(-1,2,1), {\tt h}=(2,5,-1), {\tt i}=(-2,1,1), {\tt j}=(5,1,2), {\tt k}=(-1,1,2), {\tt l}=(2,1,5), {\tt m}=(2,1,-1), {\tt n}=(5,2,-1), {\tt o}=(1,-2,1), {\tt p}=(-1,5,-2), {\tt q}=(1,1,2), {\tt r}=(-2,1,5), {\tt s}=(5,-2,-1), {\tt t}=(-1,5,2)

\begin{figure}[ht]
\center
\includegraphics[width=0.98\textwidth]{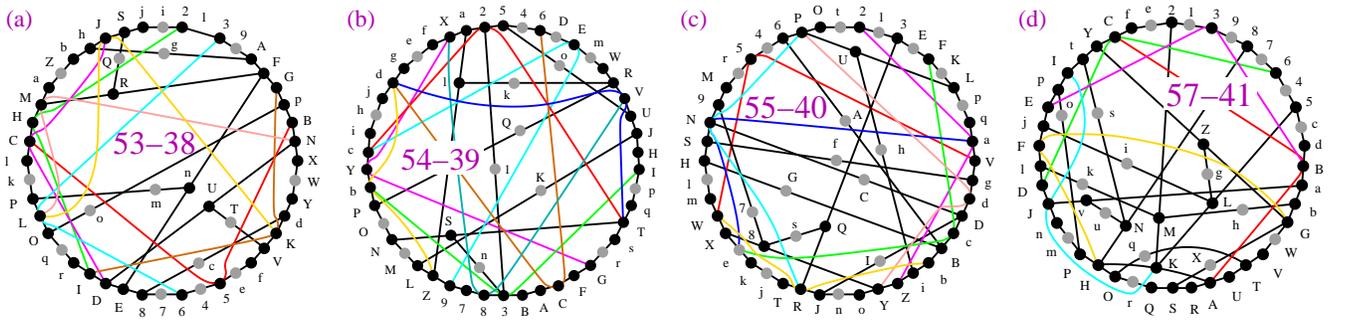}
\caption{(a-c) Critical 3D MMPHs generated by the components
  $\{0,\pm 1,\pm 2,5\}$; (a) the only 53-38; 22-gon; (b) one of the
  eight 54-39s; 23-gon; (c) the only 55-40; 22-gon; (d) the only
  57-41---the smallest MMPH generated by
  $\{0,\pm1,2,\pm5,\pm\omega,2\omega\}$; 21-gon.}
\label{fig:3d4}
\end{figure}

\subsubsection*{The smallest {\rm MMPH} generated by
  $\{0,\pm1,2,\pm5,\pm\omega,2\omega\}$, shown in
  Fig.~\ref{fig:3d4}(d)}

{\bf 57-41} {\tt 213,398,876,645,5cd,dBa,abG,GWV,VTU,UAR,RSQ,QrO,OHP,Pmn,nJD,DlF,FjE,EpI,ItY,YCf,fe2,ABC,DC6,\break EB3,FGH,IJK,LK5,MK2,NH8,KHA,XSG,YZG,gZL,hbM,ijL,klM,opP,qTO,stN,uvN,vaJ.} {\tt 1}=(1,2,5); {\tt 2}=(2,-1,0); \break{\tt 3}=(1,2,-1); {\tt 4}=(-1,2,5); {\tt 5}=(2,1,0); {\tt 6}=(-1,2,-1); {\tt 7}=(5,2,-1); {\tt 8}=(0,1,2); {\tt 9}=(-5,2,-1); {\tt A}=(0,1,0); {\tt B}=(1,0,1); {\tt C}=(1,0,-1); {\tt D}=(1,1,1); {\tt E}=(-1,1,1); {\tt F}=(0,1,-1); {\tt G}=(0,1,1); {\tt H}=(1,0,0); {\tt I}=(1,1,0); {\tt J}=(1,-1,0); {\tt K}=(0,0,1); {\tt L}=(-1,2,0); {\tt M}=(1,2,0); {\tt N}=(0,2,-1); {\tt O}=(0,-1,2); {\tt P}=(0,2,1); {\tt Q}=(-1,2$\omega$,$\omega$); {\tt R}=(1,0,$\omega$); {\tt S}=(1,$\omega$,-$\omega$);\break {\tt T}=(1,2$\omega$,$\omega$); {\tt U}=(1,0,-$\omega$); {\tt V}=(1,-$\omega$,$\omega$); {\tt W}=(2,$\omega$,-$\omega$); {\tt X}=(2,-$\omega$,$\omega$); {\tt Y}=(1,-1,1); {\tt Z}=(2,1,-1); {\tt a}=(1,1,-1); {\tt b}=(2,-1,1);\break {\tt c}=(-1,2,-5); {\tt d}=(-1,2,1); {\tt e}=(1,2,-5); {\tt f}=(1,2,1); {\tt g}=(2,1,5); {\tt h}=(2,-1,-5); {\tt i}=(2,1,-5); {\tt j}=(2,1,1); {\tt k}=(2,-1,5);\break  {\tt l}=(2,-1,-1); {\tt m}=(5,-1,2); {\tt n}=(-1,-1,2); {\tt o}=(-5,-1,2); {\tt p}=(1,-1,2); {\tt q}=(-5,2$\omega$,$\omega$); {\tt r}=(5,2$\omega$,$\omega$); {\tt s}=(5,1,2); {\tt t}=(-1,1,2); {\tt u}=(-5,1,2); {\tt v}=(1,1,2).

\end{widetext}

\subsubsection*{Original {\rm KS} set}

\parindent=12pt
The original 192-118 KS set found by Kochen and Specker
\cite{koch-speck} was not accompanied by an explicit
coordinatization. Instead, they give the equation
$\sin(\pi/10)=xy/\sqrt{(1+x^2+x^2y^2)(1+y^2+x^2y^2)}$
where $x$ and $y$, provided they satisfy this equation, are
presumably arbitrary parameters that are used to define the
vectors in their starting subgraph called $\Gamma_1$. The final
hypergraph (called $\Gamma_2$), shown in Fig.~3, consists
of 15 copies of $\Gamma_1$, with assumed---but so far
undefined---vectors rotated (using rotation
matrices) so that the copies align correctly.
Notice that Kochen and Specker in their original graphical
representation of $\Gamma_2$ dropped all 75 gray vertices thus
arriving and 117-118 hypergraph. But as we explained in
\cite[Sec.~XII]{pavicic-pra-17} if one dropped all gray vertices,
the remaining hypergraph would not be a KS set any more.

The $x,y$ satisfying the equation is not sufficient
to provide a coordinatization. E.g., one solution is
$x$=$y$=$\sqrt{\varphi}$, where $\varphi$=$(1+\sqrt{5})/2$ is the
golden ratio. When we compute vectors for just the 117 black
vertices on hypergraph $\Gamma_2$, only 115 of them are unique,
so we cannot use it even for the truncated non-critical
\cite{pavicic-entropy-19} 117-118 set. 
The complete 192-118 KS MMPH requires  $192-117=75$ additional
unique vectors (each computed as the cross product of the other two
vectors on the hyperedge), placing an even more severe constraint on
the possible values of $x$ and $y$.

We found that $x$=$1/\varphi$ and $y$=$\sqrt{\varphi}$ yield 192
unique vectors, and we used these values to compute the
coordinatization. To show the final result, let us define
$f(p,q,r)=\sqrt{2^p\sqrt{5}^q\varphi^r}$. Then we define 16
constants as follows:
$c_{1}$=$f$(-1,0,0); $c_{2}$=$f$(0,0,-1);
$c_{3}$=$f$(-1,1,-3);\hfill\break
$c_{4}$=$f$(-1,0,-2);\hfil$c_{5}$=$f$(0,-1,-6);\hfil$c_{6}$=$f$(-1,-1,-5);\hfil$c_{7}$=$f$(2,-1,0);\break
$c_{8}$=$f$(-1,-1,-3); $c_{9}$=$f$(0,-1,-2); $c_{10}$=$f$(0,0,-5); $c_{11}$\hfill\break =$f$(-1,0,-6);\hfil$c_{12}$=$f$(2,-1,-2);\hfil$c_{13}$=$f$(-1,-1,3);\hfil$c_{14}$=$f$(2,-1,-4);\break
$c_{15}$=$f(\ln(9/2-9/\sqrt{5})/\ln2,0,0)$=$3\sqrt{(5-2\sqrt{5})/10}$;
\hfill\break $c_{16}$=$f(\ln(17/4-31/(4\sqrt{5}))/\ln2,0,0)$=$\frac{1}{2}\sqrt{(85-31\sqrt{5})/5}$.

\bigskip\bigskip\bigskip

\begin{figure}[ht]
\begin{center}
\includegraphics[width=0.4\textwidth]{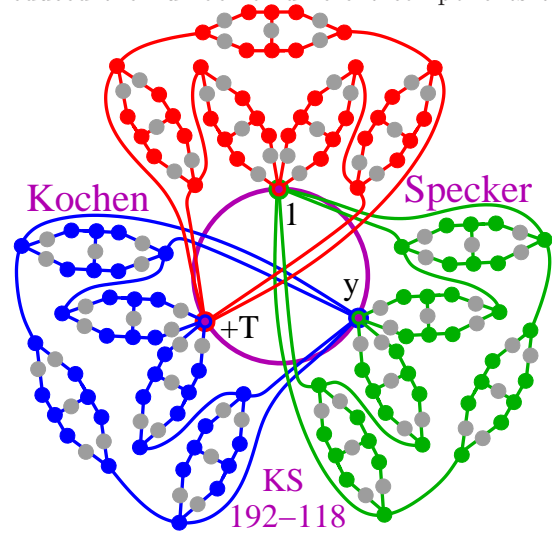}
\caption{KS MMPH of the original Kochen-Specker 192-118 set
  presented according to \cite[Fig.~6]{pmmm05a-corr}; note that
  the figure in \cite[Fig.~6]{pmmm05a-corr} and the one-to-one
  correspondence to \cite[$\Gamma_2$]{koch-speck} presented
  there is obtained exclusively from the ASCII string of the MMPH
  without any reference to its coordinatization; note also that
  one can use the string below to assign its vertices to the
  vertices of the figure by hand.}
\end{center}
\label{fig:ksorig}
\end{figure}

When normalized, the 192 vectors have 73 different
components. By using unnormalized vectors, i.e. rays, we reduced
the number of different components to 24, whose values are
$0,1,\pm c_{1}$, $\pm c_{2}$, $c_{3}$, $c_{4}$, $-c_{5}$, $\pm c_{6}$,
$c_{7}$, $\pm c_{8}$, $\pm c_{9}$, $-c_{10}$, $\pm c_{11}$, $-c_{12}$,
$-c_{13}$, $-c_{14}$, $-c_{15}$, and $-c_{16}$.
The coordinatization is presented below.

\begin{widetext}

\medskip
{\parindent=0pt

{\bf 192-118}\ \qquad \
{\tt 123,345,567,789,9AB,BC1,DEF,FGH,HIJ,JKL,LMN,NOD,PQR,RST,TUV,VWX,XYZ,ZaP,bcd,def,fgh,\break hij,jkl,lmb,1no,opq,qrs,stu,uvw,wx1,yz!,!"\#,\#\$\%,\%\&','(),)*y,-/:,:;\textless,\textless=\textgreater,\textgreater?@,@[\textbackslash,\textbackslash]-,\textasciicircum\_`,`\{|,|\}$\sim$,\break$\sim$+1+2,+2+3+4,+4+5\textasciicircum,+6+7+8,+8+9+A,+A+B+C,+C+D+E,+E+F+G,+G+H+6,y+I+J,+J+K+L,+L+M+N,+N+O+P,+P+Q+R,\break+R+Sy,+T+U+V,+V+W+X,+X+Y+Z,+Z+a+b,+b+c+d,+d+e+T,+f+g+h,+h+i+j,+j+k+l,+l+m+n,+n+o+p,+p+q+f,+r+s+t,\break+t+u+v,+v+w+x,+x+y+z,+z+!+",+"+\#+r,+\$+\%+\&,+\&+'+(,+(+)+*,+*+-+/,+/+:+;,+;+\textless+\$,+T+=+\textgreater,+\textgreater+?+@,\break+@+[+\textbackslash,+\textbackslash+]+\textasciicircum,+\textasciicircum+\_+`,+`+\{+T,4A+|,GM+\},SY+$\sim$,ek++1,pv++2,"(++3,;[++4,\{+3++5,+9+F++6,+K+Q++7,\break+W+c++8,+i+o++9,+u+!++A,+'+:++B,+?+\_++C,D7y,PJy,bVy,shy,-\%+T,\textasciicircum\textgreater+T,$\sim$+C+T,+N+6+T,+f+Z1,+r+l1,\break+\$+x1,+\textbackslash+*1,1y+T.}

\smallskip
{\tt 1{\rm =}(1,0,0); 2{\rm =}(0,{\rm -}$c_{2}$,$c_{4}$); 3{\rm =}(0,$c_{4}$,$c_{2}$); 4{\rm =}($c_{8}$,{\rm -}$c_{10}$,$c_{11}$); 5{\rm =}($c_{1}$,$c_{9}$,{\rm -}$c_{8}$); 6{\rm =}({\rm -}$c_{1}$,$c_{7}$,$c_{8}$); 7{\rm =}($c_{8}$,0,$c_{1}$); \break8{\rm =}({\rm -}$c_{1}$,{\rm -}$c_{12}$,$c_{8}$);$\>$9{\rm =}($c_{4}$,{\rm -}$c_{9}$,{\rm -}$c_{6}$);$\>$A{\rm =}($c_{4}$,$c_{9}$,$c_{6}$);$\>$B{\rm =}(0,$c_{6}$,{\rm -}$c_{9}$);$\>$C{\rm =}(0,$c_{9}$,$c_{6}$);$\>$D{\rm =}($c_{1}$,0,{\rm -}$c_{8}$);$\>$E{\rm =}($c_{8}$,{\rm -}$c_{12}$,$c_{1}$);\break
  F{\rm =}($c_{6}$,$c_{9}$,$c_{4}$); G{\rm =}($c_{1}$,{\rm -}$c_{2}$,$c_{3}$); H{\rm =}({\rm -}$c_{13}$,{\rm -}$c_{9}$,$c_{1}$); I{\rm =}($c_{8}$,{\rm -}$c_{12}$,{\rm -}$c_{11}$); J{\rm =}($c_{3}$,0,$c_{1}$); K{\rm =}({\rm -}$c_{8}$,{\rm -}$c_{14}$,$c_{11}$);\break L{\rm =}($c_{8}$,{\rm -}$c_{5}$,{\rm -}$c_{11}$);$\>$M{\rm =}($c_{1}$,$c_{2}$,$c_{8}$);$\>$N{\rm =}($c_{8}$,{\rm -}$c_{9}$,$c_{1}$);$\>$O{\rm =}($c_{8}$,$c_{7}$,$c_{1}$);$\>$P{\rm =}({\rm -}$c_{13}$,0,$c_{1}$);$\>$Q{\rm =}({\rm -}$c_{11}$,{\rm -}$c_{12}$,{\rm -}$c_{8}$);$\>$R{\rm =}({\rm -}$c_{1}$,$c_{9}$,{\rm -}$c_{13}$);\break S{\rm =}($c_{4}$,$c_{9}$,{\rm -}$c_{6}$);$\>$T{\rm =}($c_{3}$,{\rm -}$c_{2}$,{\rm -}$c_{1}$);$\>$U{\rm =}($c_{1}$,$c_{7}$,{\rm -}$c_{13}$);$\>$V{\rm =}($c_{1}$,0,$c_{3}$);$\>$W{\rm =}($c_{11}$,{\rm -}$c_{12}$,{\rm -}$c_{8}$);$\>$X{\rm =}($c_{1}$,$c_{9}$,{\rm -}$c_{13}$);$\>$Y{\rm =}($c_{4}$,{\rm -}$c_{9}$,$c_{6}$);\break Z{\rm =}($c_{3}$,$c_{2}$,$c_{1}$); a{\rm =}({\rm -}$c_{1}$,$c_{7}$,{\rm -}$c_{13}$); b{\rm =}($c_{1}$,0,{\rm -}$c_{13}$); c{\rm =}($c_{8}$,{\rm -}$c_{14}$,$c_{11}$); d{\rm =}({\rm -}$c_{8}$,{\rm -}$c_{5}$,{\rm -}$c_{11}$); e{\rm =}($c_{1}$,{\rm -}$c_{2}$,{\rm -}$c_{8}$);\break
  f{\rm =}({\rm -}$c_{8}$,{\rm -}$c_{9}$,$c_{1}$);$\>$g{\rm =}({\rm -}$c_{8}$,$c_{7}$,$c_{1}$);$\>$h{\rm =}($c_{1}$,0,$c_{8}$);$\>$i{\rm =}({\rm -}$c_{8}$,{\rm -}$c_{12}$,$c_{1}$);$\>$j{\rm =}({\rm -}$c_{6}$,$c_{9}$,$c_{4}$);$\>$k{\rm =}({\rm -}$c_{1}$,{\rm -}$c_{2}$,$c_{3}$);$\>$l{\rm =}({\rm -}$c_{13}$,$c_{9}$,{\rm -}$c_{1}$);\break m{\rm =}({\rm -}$c_{8}$,{\rm -}$c_{12}$,{\rm -}$c_{11}$);$\>$n{\rm =}(0,$c_{2}$,$c_{4}$);$\>$o{\rm =}(0,$c_{4}$,{\rm -}$c_{2}$);$\>$p{\rm =}($c_{8}$,{\rm -}$c_{10}$,{\rm -}$c_{11}$);$\>$q{\rm =}($c_{1}$,$c_{9}$,$c_{8}$);$\>$r{\rm =}({\rm -}$c_{1}$,$c_{7}$,{\rm -}$c_{8}$);$\>$s{\rm =}({\rm -}$c_{8}$,0,$c_{1}$);\break t{\rm =}($c_{4}$,$c_{7}$,$c_{6}$); u{\rm =}($c_{4}$,{\rm -}$c_{5}$,$c_{6}$); v{\rm =}($c_{3}$,$c_{2}$,{\rm -}$c_{1}$); w{\rm =}(0,$c_{1}$,$c_{2}$); x{\rm =}(0,{\rm -}$c_{2}$,$c_{1}$); y{\rm =}(0,1,0); z{\rm =}($c_{6}$,0,$c_{9}$);\break
  !{\rm =}($c_{9}$,0,{\rm -}$c_{6}$);$\>$"{\rm =}($c_{6}$,$c_{4}$,$c_{9}$);$\>$\#{\rm =}({\rm -}$c_{6}$,$c_{4}$,{\rm -}$c_{9}$);$\>$\${\rm =}($c_{8}$,{\rm -}$c_{1}$,{\rm -}$c_{12}$);$\>$\%{\rm =}($c_{1}$,$c_{8}$,0);$\>$\&{\rm =}($c_{8}$,{\rm -}$c_{1}$,$c_{7}$);$\>$'{\rm =}({\rm -}$c_{8}$,$c_{1}$,$c_{9}$);\break({\rm =}($c_{11}$,$c_{8}$,{\rm -}$c_{10}$);$\>$){\rm =}($c_{2}$,0,$c_{4}$);$\>$*{\rm =}($c_{4}$,0,{\rm -}$c_{2}$);$\>$-{\rm =}({\rm -}$c_{8}$,$c_{1}$,0);$\>$/{\rm =}($c_{1}$,$c_{8}$,{\rm -}$c_{12}$);$\>$:{\rm =}($c_{4}$,$c_{6}$,$c_{9}$);$\>$ ;{\rm =}($c_{3}$,$c_{1}$,{\rm -}$c_{2}$);\break\textless{\rm =}($c_{1}$,{\rm -}$c_{13}$,{\rm -}$c_{9}$); ={\rm =}({\rm -}$c_{11}$,$c_{8}$,{\rm -}$c_{12}$); \textgreater{\rm =}($c_{1}$,$c_{3}$,0); ?{\rm =}($c_{11}$,{\rm -}$c_{8}$,{\rm -}$c_{14}$); @{\rm =}({\rm -}$c_{11}$,$c_{8}$,{\rm -}$c_{5}$); [{\rm =}($c_{8}$,$c_{1}$,$c_{2}$);\break
\textbackslash{\rm =}($c_{1}$,$c_{8}$,{\rm -}$c_{9}$); ]{\rm =}($c_{1}$,
$c_{8}$,$c_{7}$); \textasciicircum{\rm =}($c_{1}$,{\rm -}$c_{13}$,0); \_{\rm =}($c_{8}$,$c_{11}$,{\rm -}$c_{14}$); 
`{\rm =}({\rm -}$c_{8}$,{\rm -}$c_{11}$,{\rm -}$c_{5}$); \{{\rm =}(0,{\rm -}$c_{9}$,$c_{4}$);\break|{\rm =}($c_{8}$,{\rm -}$c_{11}$,{\rm -}$c_{5}$); \}{\rm =}({\rm -}$c_{8}$,$c_{11}$,{\rm -}$c_{14}$); $\sim${\rm =}($c_{3}$,$c_{1}$,0); +1{\rm =}({\rm -}$c_{13}$,$c_{1}$,$c_{7}$); +2{\rm =}({\rm -}$c_{1}$,$c_{3}$,{\rm -}$c_{2}$); +3{\rm =}({\rm -}$c_{6}$,$c_{4}$,$c_{9}$);\break
+4{\rm =}({\rm -}$c_{13}$,{\rm -}$c_{1}$,$c_{9}$); +5{\rm =}({\rm -}$c_{8}$,{\rm -}$c_{11}$,{\rm -}$c_{12}$); +6{\rm =}($c_{8}$,$c_{1}$,0); +7{\rm =}($c_{1}$,{\rm -}$c_{8}$,{\rm -}$c_{12}$); +8{\rm =}($c_{4}$,{\rm -}$c_{6}$,$c_{9}$); 
+9{\rm =}($c_{3}$,{\rm -}$c_{1}$,{\rm -}$c_{2}$);\break+A{\rm =}({\rm -}$c_{1}$,{\rm -}$c_{13}$,$c_{9}$);$\>$+B{\rm =}({\rm -}$c_{11}$,{\rm -}$c_{8}$,{\rm -}$c_{12}$);$\>$+C{\rm =}({\rm -}$c_{13}$,$c_{1}$,0);$\>$+D{\rm =}($c_{11}$,$c_{8}$,{\rm -}$c_{14}$);$\>$+E{\rm =}({\rm -}$c_{11}$,{\rm -}$c_{8}$,{\rm -}$c_{5}$);$\>$+F{\rm =}({\rm -}$c_{8}$,$c_{1}$,{\rm -}$c_{2}$);\break+G{\rm =}($c_{1}$,{\rm -}$c_{8}$,{\rm -}$c_{9}$); +H{\rm =}($c_{1}$,{\rm -}$c_{8}$,$c_{7}$); +I{\rm =}($c_{1}$,0,{\rm -}$c_{2}$); +J{\rm =}($c_{2}$,0,$c_{1}$); +K{\rm =}({\rm -}$c_{1}$,$c_{3}$,$c_{2}$); +L{\rm =}($c_{6}$,$c_{4}$,{\rm -}$c_{5}$);\break+M{\rm =}($c_{6}$,$c_{4}$,$c_{7}$); +N{\rm =}($c_{1}$,{\rm -}$c_{8}$,0); +O{\rm =}({\rm -}$c_{8}$,{\rm -}$c_{1}$,$c_{7}$); +P{\rm =}($c_{8}$,$c_{1}$,$c_{9}$); +Q{\rm =}({\rm -}$c_{11}$,$c_{8}$,{\rm -}$c_{10}$); +R{\rm =}({\rm -}$c_{2}$,0,$c_{4}$);\break
+S{\rm =}($c_{4}$,0,$c_{2}$);$\>$+T{\rm =}(0,0,1);$\>$+U{\rm =}({\rm -}$c_{2}$,$c_{1}$,0);$\>$+V{\rm =}($c_{1}$,$c_{2}$,0);$\>$+W{\rm =}($c_{2}$,{\rm -}$c_{1}$,$c_{3}$);$\>$+X{\rm =}({\rm -}$c_{5}$,$c_{6}$,$c_{4}$);$\>$+Y{\rm =}($c_{7}$,$c_{6}$,$c_{4}$);\break+Z{\rm =}(0,$c_{1}$,{\rm -}$c_{8}$);$\>$+a{\rm =}($c_{7}$,{\rm -}$c_{8}$,{\rm -}$c_{1}$);$\>$+b{\rm =}($c_{9}$,$c_{8}$,$c_{1}$);$\>$+c{\rm =}({\rm -}$c_{10}$,{\rm -}$c_{11}$,$c_{8}$);$\>$+d{\rm =}($c_{4}$,{\rm -}$c_{2}$,0);$\>$+e{\rm =}($c_{2}$,$c_{4}$,0);\break+f{\rm =}(0,$c_{8}$,$c_{1}$);$\>$+g{\rm =}({\rm -}$c_{12}$,$c_{1}$,{\rm -}$c_{8}$); +h{\rm =}($c_{9}$,$c_{4}$,
{\rm -}$c_{6}$); +i{\rm =}({\rm -}$c_{2}$,$c_{3}$,{\rm -}$c_{1}$); +j{\rm =}($c_{9}$,{\rm -}$c_{1}$,{\rm -}$c_{13}$); +k{\rm =}({\rm -}$c_{12}$,{\rm -}$c_{11}$,{\rm -}$c_{8}$);\break+l{\rm =}(0,{\rm -}$c_{13}$,$c_{1}$); +m{\rm =}({\rm -}$c_{14}$,
$c_{11}$,$c_{8}$); +n{\rm =}({\rm -}$c_{5}$,{\rm -}$c_{11}$,{\rm -}$c_{8}$); +o{\rm =}({\rm -}$c_{2}$,{\rm -}$c_{8}$,$c_{1}$); +p{\rm =}({\rm -}$c_{9}$,$c_{1}$,{\rm -}$c_{8}$); +q{\rm =}($c_{7}$,$c_{1}$,{\rm -}$c_{8}$);\break
+r{\rm =}(0,$c_{3}$,$c_{1}$); +s{\rm =}($c_{7}$,{\rm -}$c_{13}$,$c_{1}$); +t{\rm =}({\rm -}$c_{2}$,{\rm -}$c_{1}$,$c_{3}$); +u{\rm =}($c_{9}$,{\rm -}$c_{6}$,$c_{4}$); +v{\rm =}($c_{9}$,{\rm -}$c_{13}$,{\rm -}$c_{1}$); +w{\rm =}({\rm -}$c_{12}$,{\rm -}$c_{8}$,{\rm -}$c_{11}$);\break+x{\rm =}(0,$c_{1}$,{\rm -}$c_{13}$);$\>$+y{\rm =}({\rm -}$c_{14}$,$c_{8}$,$c_{11}$);$\>$+z{\rm =}({\rm -}$c_{5}$,{\rm -}$c_{8}$,{\rm -}$c_{11}$);$\>$+!{\rm =}($c_{4}$,0,{\rm -}$c_{9}$);$\>$+"{\rm =}({\rm -}$c_{5}$,$c_{8}$,$\>${\rm -}$c_{11}$);$\>$+\#{\rm =}({\rm -}$c_{14}$,{\rm -}$c_{8}$,$c_{11}$);\break
+\${\rm =}(0,$c_{1}$,$c_{3}$); +\%{\rm =}({\rm -}$c_{12}$,{\rm -}$c_{11}$,$c_{8}$); +\&{\rm =}({\rm -}$c_{9}$,$c_{1}$,{\rm -}$c_{13}$); +'{\rm =}({\rm -}$c_{2}$,$c_{3}$,$c_{1}$); +({\rm =}($c_{9}$,$c_{4}$,$c_{6}$); +){\rm =}({\rm -}$c_{12}$,$c_{1}$,$c_{8}$);\break+*{\rm =}(0,{\rm -}$c_{8}$,$c_{1}$); +-{\rm =}($c_{7}$,$c_{1}$,$c_{8}$); +/{\rm =}({\rm -}$c_{9}$,$c_{1}$,$c_{8}$); +:{\rm =}($c_{2}$,$c_{8}$,$c_{1}$); +;{\rm =}({\rm -}$c_{5}$,{\rm -}$c_{11}$,$c_{8}$); +\textless{\rm =}({\rm -}$c_{14}$,$c_{11}$,{\rm -}$c_{8}$);\break+={\rm =}($c_{9}$,$c_{6}$,0); +\textgreater{\rm =}($c_{6}$,{\rm -}$c_{9}$,0); +?{\rm =}($c_{9}$,$c_{6}$,$c_{4}$); +@{\rm =}({\rm -}$c_{9}$,{\rm -}$c_{6}$,$c_{4}$); +[{\rm =}({\rm -}$c_{12}$,$c_{8}$,{\rm -}$c_{1}$); +\textbackslash{\rm =}(0,$c_{1}$,$c_{8}$);\break
+]{\rm =}($c_{7}$,$c_{8}$,{\rm -}$c_{1}$); +\textasciicircum{\rm =}($c_{9}$,{\rm -}$c_{8}$,$c_{1}$); +\_{\rm =}({\rm -}$c_{10}$,$c_{11}$,$c_{8}$); +`{\rm =}($c_{4}$,$c_{2}$,0); +\{{\rm =}({\rm -}$c_{2}$,$c_{4}$,0); +|{\rm =}($c_{4}$,{\rm -}$c_{5}$,{\rm -}$c_{16}$);\break+\}{\rm =}({\rm -}$c_{9}$,$c_{6}$,$c_{2}$); +$\sim${\rm =}(0,$c_{6}$,$c_{9}$); ++1{\rm =}($c_{9}$,$c_{6}$,$c_{2}$); ++2{\rm =}({\rm -}$c_{1}$,{\rm -}$c_{5}$,{\rm -}$c_{15}$); ++3{\rm =}({\rm -}$c_{16}$,$c_{4}$,{\rm -}$c_{5}$); ++4{\rm =}($c_{2}$,{\rm -}$c_{9}$,$c_{6}$);\break++5{\rm =}({\rm -}$c_{13}$,{\rm -}$c_{11}$,{\rm -}$c_{5}$); ++6{\rm =}($c_{2}$,$c_{9}$,$c_{6}$); ++7{\rm =}({\rm -}$c_{15}$,{\rm -}$c_{1}$,{\rm -}$c_{5}$); ++8{\rm =}({\rm -}$c_{5}$,{\rm -}$c_{15}$,
{\rm -}$c_{1}$); ++9{\rm =}($c_{6}$,$c_{2}$,$c_{9}$);\break++A{\rm =}({\rm -}$c_{5}$,{\rm -}$c_{13}$,{\rm -}$c_{11}$); ++B{\rm =}($c_{6}$,$c_{2}$,{\rm -}$c_{9}$); ++C{\rm =}({\rm -}$c_{5}$,{\rm -}$c_{16}$,$c_{4}$).}

\medskip

The components may be used to generate a master MMPH whose class
contains the MMPH 192-118. The master has 2416 vertices and 1432
hyperedges.

\bigskip

\subsection*{5-dim}
{\bf 29-16}\qquad{\tt HOINJ,JGSTF,FT4Q8,85679,92LOH,PQRST,KLMNO,CDEIO,ABEGT,34DMO,12BRT,237CO,146AT,279OP,\break 468KT,5EJOT.} {\tt 1}=(1,-1,1,0,-1); {\tt 2}=(1,0,-1,0,0); {\tt 3}=(1,-1,1,1,0); {\tt 4}=(0,1,1,0,0); {\tt 5}=(0,0,1,0,0); {\tt 6}=(1,0,0,0,1); {\tt 7}=(0,1,0,1,0); {\tt 8}=(1,0,0,0,-1); {\tt 9}=(0,1,0,-1,0); {\tt A}=(1,1,-1,0,-1); {\tt B}=(1,1,1,0,1); {\tt C}=(1,1,1,-1,0); {\tt D}=(1,1,-1,1,0);\break  {\tt E}=(1,-1,0,0,0); {\tt F}=(1,-1,1,0,1); {\tt G}=(0,0,1,0,-1); {\tt H}=(1,-1,1,-1,0); {\tt I}=(0,0,1,1,0); {\tt J}=(1,1,0,0,0); {\tt K}=(0,1,-1,0,0); {\tt L}=(1,1,1,1,0); {\tt M}=(1,0,0,-1,0); {\tt N}=(1,-1,-1,1,0); {\tt O}=(0,0,0,0,1); {\tt P}=(1,0,1,0,0); {\tt Q}=(1,1,-1,0,1); {\tt R}=(0,1,0,0,-1);\break {\tt S}=(-1,1,1,0,1); {\tt T}=(0,0,0,1,0).

\medskip
{\bf 31-17} {\tt 12345,16789,16ABC,13DEB,14D8F,1GHF9,1IJ8K,1IHAE,LMGI5,L27NO,L2PQC,M27RS,2TJUO,2THPR,23VUN,\break 24VPS,2PQRS.} {\tt 1}=(0,0,0,0,1); {\tt 2}=(0,0,0,1,0); {\tt 3}=(1,1,0,0,0); {\tt 4}=(1,-1,0,0,0); {\tt 5}=(0,0,1,0,0); {\tt 6}=(1,0,0,-1,0); {\tt 7}=(0,1,1,0,0); {\tt 8}=(1,1,-1,1,0); {\tt 9}=(1,-1,1,1,0); {\tt A}=(1,1,1,1,0); {\tt B}=(1,-1,-1,1,0); {\tt C}=(0,1,-1,0,0); {\tt D}=(0,0,1,1,0); {\tt E}=(1,-1,1,-1,0);\break  {\tt F}=(1,1,1,-1,0); {\tt G}=(0,1,0,1,0); {\tt H}=(1,0,-1,0,0); {\tt I}=(0,1,0,-1,0); {\tt J}=(1,0,1,0,0); {\tt K}=(-1,1,1,1,0); {\tt L}=(1,0,0,0,1);\break {\tt M}=(1,0,0,0,-1); {\tt N}=(1,-1,1,0,-1); {\tt O}=(1,1,-1,0,-1); {\tt P}=(1,1,1,0,-1); {\tt Q}=(-1,1,1,0,1); {\tt R}=(1,-1,1,0,1); {\tt S}=(1,1,-1,0,1); {\tt T}=(0,1,0,0,1); {\tt U}=(1,-1,-1,0,1); {\tt V}=(0,0,1,0,1).

\subsection*{7-dim}

{\bf 13-4} {\tt 1234567,456789A,123ABCD,4567BCD.}

\medskip
{\bf 34-14} {\tt 1234567, 189A5BC, 189DE7F, 189GHIJ, 189KHBL, 2MNDOIP, 2MNEOCL, 2MNGK6F, QRNSAJP, QT4U567, RTV9567, WXMS567, WYV8AJP, XY3U567.} {\tt 1}=(0,0,0,1,0,0,0); {\tt 2}=(0,0,1,0,0,0,0); {\tt 3}=(1,-1,0,0,0,0,0); {\tt 4}=(1,1,0,0,0,0,0); {\tt 5}=(0,0,0,0,0,0,1); {\tt 6}=(0,0,0,0,1,1,0); {\tt 7}=(0,0,0,0,1,-1,0); {\tt 8}=(0,1,-1,0,0,0,0); {\tt 9}=(0,1,1,0,0,0,0); {\tt A}=(0,0,0,0,1,0,0); {\tt B}=(1,0,0,0,0,-1,0); {\tt C}=(1,0,0,0,0,1,0); {\tt D}=(1,0,0,0,1,1,-1); {\tt E}=(-1,0,0,0,1,1,1); {\tt F}=(1,0,0,0,0,0,1); {\tt G}=(1,0,0,0,1,-1,-1); {\tt H}=(1,0,0,0,1,1,1); {\tt I}=(1,0,0,0,-1,0,0); {\tt J}=(0,0,0,0,0,1,-1); {\tt K}=(1,0,0,0,-1,1,-1); {\tt L}=(0,0,0,0,1,0,-1); {\tt M}=(0,1,0,1,0,0,0); {\tt N}=(0,1,0,-1,0,0,0); {\tt O}=(1,0,0,0,1,-1,1); {\tt P}=(0,0,0,0,0,1,1); {\tt Q}=(-1,1,1,1,0,0,0); {\tt R}=(1,1,-1,1,0,0,0); {\tt S}=(1,0,1,0,0,0,0); {\tt T}=(1,-1,1,1,0,0,0); {\tt U}=(0,0,1,-1,0,0,0); {\tt V}=(1,0,0,-1,0,0,0); {\tt W}=(1,-1,-1,1,0,0,0); {\tt X}=(1,1,-1,-1,0,0,0); {\tt Y}=(1,1,1,1,0,0,0).

\medskip
{\bf 28(58)-14} {\tt 3451678, 4LF8BJK, PHG7QRS, 349A16(T), 38BCD(UV), F8QJO(WX), 4HIJK(YZ), AMNCD(ab), 3452(cde), EFG2(fgh), 9MNO(ijk), PLRS(lmn), EI7(opqr), 12(stuvw).} {\tt 1}=(0,1,0,0,0,0,0); {\tt 2}=(1,0,0,0,0,0,0); {\tt 3}=(0,0,1,1,1,-1,0);  {\tt 4}=(0,0,1,1,-1,1,0); {\tt 5}=(0,0,1,-1,0,0,0); {\tt 6}=(1,0,0,0,0,0,1); {\tt 7}=(1,0,0,0,1,1,-1); {\tt 8}=(-1,0,0,0,1,1,1); {\tt 9}=(0,0,1,-1,1,1,0);  {\tt A}=(0,0,-1,1,1,1,0); {\tt B}=(1,0,-1,1,0,0,1); {\tt C}=(1,-1,0,-1,1,0,0); {\tt D}=(1,1,1,0,0,1,0); {\tt E}=(0,1,0,-1,-1,1,0); {\tt F}=(0,1,1,0,0,-1,1);  {\tt G}=(0,-1,1,0,0,1,1); {\tt H}=(0,1,0,1,1,0,1); {\tt I}=(0,1,1,0,0,-1,-1); {\tt J}=(1,-1,1,0,1,0,0); {\tt K}=(1,1,0,-1,0,1,0); {\tt L}=(0,1,0,1,1,0,-1);  {\tt M}=(0,1,0,0,1,-1,-1); {\tt N}=(0,1,-1,-1,0,0,1); {\tt O}=(1,1,0,1,0,1,0); {\tt Q}=(1,0,-1,-1,0,0,1); {\tt P}=(0,0,1,-1,1,-1,0); {\tt R}=(1,1,1,0,-1,0,0);  {\tt S}=(1,-1,0,1,0,-1,0); {\tt T}=(1,0,0,0,0,0,-1); {\tt U}=(0,1,0,-1,0,-1,1); {\tt V}=(0,1,-1,0,1,0,-1); {\tt W}=(0,0,1,-1,-1,1,0); {\tt X}=(0,1,0,-1,1,0,-1);  {\tt Y}=(1,0,-1,1,0,0,-1); {\tt Z}=(1,0,0,0,-1,-1,1); {\tt a}=(1,0,0,1,0,-1,1); {\tt b}=(-1,0,1,0,1,0,1); {\tt c}=(0,1,0,0,0,0,1); {\tt d}=(0,1,0,0,1,1,-1);  {\tt e}=(0,-1,0,0,1,1,1); {\tt f}=(0,0,1,1,-1,0,-1); {\tt g}=(0,0,1,-1,1,0,-1); {\tt h}=(0,1,0,1,1,1,0); {\tt i}=(1,-1,-1,0,1,0,0); {\tt j}=(-1,0,0,1,1,0,1);  {\tt k}=(1,0,1,0,0,-1,1); {\tt l}=(0,1,-1,0,0,-1,1); {\tt m}=(1,0,0,0,1,1,1); {\tt n}=(-1,0,1,1,0,0,1); {\tt o}=(0,0,1,-1,1,0,1); {\tt p}=(1,-1,1,0,-1,0,0);  {\tt q}=(-1,0,1,1,0,1,0); {\tt r}=(1,1,0,0,1,0,-1); {\tt s}=(0,0,0,1,1,1,-1); {\tt t}=(0,0,0,1,1,-1,1); {\tt u}=(0,0,0,1,-1,1,1); {\tt v}=(0,0,0,-1,1,1,1); {\tt w}=(0,0,1,0,0,0,0).

\medskip
{\bf 207-97}\quad {\tt 1234567, 89ABCDE, 8FGHIJK, 8LMNOPE, QRS6TUV, QRS6WXY, QRSZab7, QRScde7, QRfg6TW, QRfgZc7, Qf6hijk, QglLmn7, Qaopqrs, QetuvOw, Qxyzq!", Q\#\$V\%\&', Q(yuU)*, QLnX-/r, :;6hN\textless =, :\textgreater ?@[\textbackslash ], :\textasciicircum \_`\{]$\vert$, \}$\sim$\textasciicircum +1M+27, \}$\sim$+3+4mG7, \}+5+6@+7+8+9, \}e+6v\textless +A+B,  \}(+C+DY\{+E, Rf6+F+GC[, RaoN+H+I+J, Rbo+K+7\textbackslash +L, R+M+NX-+O+P, R+Q+R+S+7+T!, R+U\_Y-+8+V, R+U+WGN+X+Y, R+ZMV\%+8+V,  R9+aU\%+A+I, R+b+RX-+A+I, +c+d6v+e+X+H, +c+ftuv+X+P, +cx+g+h\%+i+P, +c+Znv+H!P, +c(+Ru+j)+E,  +k+f+l+m+n+O+o, +k+p+N+S+q+r+s, +td+l+K\textless s+B, +tl+u@+v+w+x, +t+y+z+a+F+q+x,  +!+5+"+\#k'+\$, +!b+\%z+\&+'+V, +!+(+z+)+\#+X+x, +*+f+"+\#+w'+-, +*+Z+/+\#+:+r+s, +*+Z+;+D`+\textless +V, +=+1\_U+\textgreater +?+9, $\sim$+f+@+[+\textbackslash +]+P, $\sim$b+\textasciicircum h+\textbackslash \&+A+\_, $\sim$+Q+/+mC+`+s, f+5H+\{+GC+$\vert$,  fxyI+\}+I+J, f\#AI+\}+O+E, f(+$\sim$`++1+A++2, f(y+K+7++3P, f++4\_+j++5/++6, f+(M+h+\textgreater ++7+$\vert$,  ++8+dde++9++A7, ++8e++9+Sq+\textgreater +V, ++8e++A+[k+A', ++8+MA+K+vq++B, ++8+U++C+[j"+s, ++8+U++CN+v+w+Y, ++8+U+2U++7r+L, ++8+Zm++DO"+s, ++8+ZnX+\}r+L, ++8+(+zn+[j++B,  ++E+f+\textasciicircum ++Fi++G*, ++E+pnV++H\textbackslash ++I, ++E(+C++J++K++3+`, ++E+y++CX++L+\$$\vert$, ;+5+\textasciicircum h=\&+\_,  ;x+u++J+vq++M, ;\#+z+C+\#++N++O, +dx+gB+v[+Y, +d(+gU++7+O*, +d(+Rp+7++3+T, +dF+zGv+X+Y,  +dLMX++1/$\vert$, +dLMV++7+8+$\vert$, g+5+\{u++P+:*, g+5+\{++D+H\&', g+M+Np+vq++Q, g+M+a+h++7'++R,  g+Q+gB+v++K++Q, g+Q+R`+\}'++R, glm++D+XD+s, glnI+\}\textbackslash +L, g9+N`+\}+i*,  gF+z+2++D+X++Q, gLnN+H++3P.}

\smallskip
{\tt 1}=(0,0,0,1,1,1,-1); {\tt 2}=(0,0,0,1,1,-1,1); {\tt 3}=(0,0,0,1,-1,1,1); {\tt 4}=(0,0,0,-1,1,1,1); {\tt 5}=(0,0,1,0,0,0,0); {\tt 8}=(0,0,1,1,1,1,0); {\tt Q}=(0,0,1,1,1,-1,0); {\tt :}=(0,0,1,1,-1,0,1); {\tt \}}=(0,0,1,1,-1,0,-1); {\tt R}=(0,0,1,1,-1,1,0); {\tt +c}=(0,0,1,1,-1,-1,0); {\tt S}=(0,0,1,-1,0,0,0); {\tt +k}=(0,0,1,-1,0,1,1); {\tt +t}=(0,0,1,-1,0,1,-1); {\tt +!}=(0,0,1,-1,0,-1,1); {\tt +*}=(0,0,-1,1,0,1,1); {\tt +=}=(0,0,1,-1,1,0,1); {\tt $\sim$}=(0,0,1,-1,1,0,-1); {\tt f}=(0,0,1,-1,1,1,0); {\tt ++8}=(0,0,1,-1,1,-1,0); {\tt ++E}=(0,0,1,-1,-1,0,1); {\tt ;}=(0,0,-1,1,1,0,1); {\tt +d}=(0,0,1,-1,-1,1,0); {\tt g}=(0,0,-1,1,1,1,0); {\tt 6}=(0,1,0,0,0,0,0); {\tt Z}=(0,1,0,0,0,0,1); {\tt c}=(0,1,0,0,0,0,-1); {\tt d}=(0,1,0,0,1,1,1); {\tt a}=(0,1,0,0,1,1,-1); {\tt +f}=(0,1,0,0,1,-1,1); {\tt +5}=(0,1,0,0,1,-1,-1); {\tt \textgreater }=(0,1,0,0,-1,1,-1); {\tt e}=(0,1,0,0,-1,-1,1); {\tt b}=(0,-1,0,0,1,1,1); {\tt x}=(0,1,0,1,0,1,-1); {\tt +M}=(0,1,0,1,0,-1,1); {\tt +Q}=(0,1,0,1,0,-1,-1); {\tt +U}=(0,1,0,1,1,0,1); {\tt +Z}=(0,1,0,1,1,0,-1); {\tt \textasciicircum }=(0,1,0,1,1,1,0); {\tt +3}=(0,1,0,1,1,-1,0); {\tt l}=(0,1,0,1,-1,0,-1); {\tt +p}=(0,1,0,1,-1,1,0); {\tt 9}=(0,1,0,-1,0,1,1); {\tt +b}=(0,1,0,-1,0,1,-1); {\tt \#}=(0,1,0,-1,0,-1,1); {\tt (}=(0,-1,0,1,0,1,1); {\tt F}=(0,1,0,-1,1,0,1); {\tt L}=(0,1,0,-1,1,0,-1); {\tt +y}=(0,1,0,-1,1,-1,0); {\tt ++4}=(0,1,0,-1,-1,0,1); {\tt +(}=(0,-1,0,1,1,0,1); {\tt +1}=(0,1,0,-1,-1,1,0); {\tt +4}=(0,-1,0,1,1,1,0); {\tt m}=(0,1,1,0,0,1,1); {\tt ++C}=(0,1,1,0,0,1,-1); {\tt M}=(0,1,1,0,0,-1,1); {\tt \_}=(0,1,1,0,0,-1,-1); {\tt +g}=(0,1,1,0,1,0,1); {\tt +N}=(0,1,1,0,1,0,-1); {\tt +u}=(0,1,1,0,1,-1,0); {\tt +$\sim$}=(0,1,1,0,-1,0,1); {\tt A}=(0,1,1,0,-1,0,-1); {\tt +/}=(0,1,1,0,-1,1,0); {\tt ++9}=(0,1,1,1,0,0,-1); {\tt +@}=(0,1,1,1,0,1,0); {\tt +\%}=(0,1,1,1,1,0,0); {\tt +l}=(0,1,1,1,-1,0,0); {\tt +z}=(-1,1,1,1,0,0,0); {\tt +6}=(0,1,1,-1,0,1,0); {\tt +W}=(1,-1,-1,1,0,0,0); {\tt G}=(0,1,-1,0,0,1,-1); {\tt n}=(0,1,-1,0,0,-1,1); {\tt +2}=(0,-1,1,0,0,1,1); {\tt y}=(0,1,-1,0,1,0,1); {\tt \$}=(0,1,-1,0,1,0,-1); {\tt +)}=(0,1,-1,0,1,-1,0); {\tt +R}=(0,1,-1,0,-1,0,1); {\tt +a}=(0,-1,1,0,1,0,1); {\tt +C}=(0,1,-1,0,-1,1,0); {\tt +;}=(0,-1,1,0,1,1,0); {\tt o}=(0,1,-1,1,0,0,1); {\tt t}=(0,1,-1,1,0,0,-1); {\tt ?}=(0,1,-1,1,0,-1,0); {\tt H}=(1,-1,-1,0,1,0,0); {\tt +\{}=(0,1,-1,-1,0,0,1); {\tt ++A}=(0,-1,1,1,0,0,1); {\tt +\textasciicircum }=(0,1,-1,-1,0,1,0); {\tt u}=(-1,1,1,0,0,1,0); {\tt +"}=(0,-1,1,1,1,0,0); {\tt +D}=(-1,1,1,0,0,0,1); {\tt ++F}=(1,-1,-1,0,0,0,1); {\tt 7}=(1,0,0,0,0,0,0); {\tt T}=(1,0,0,0,0,0,1); {\tt W}=(1,0,0,0,0,0,-1); {\tt X}=(1,0,0,0,1,1,1); {\tt U}=(1,0,0,0,1,1,-1); {\tt `}=(1,0,0,0,1,-1,1); {\tt +h}=(1,0,0,0,1,-1,-1); {\tt I}=(1,0,0,0,-1,1,1); {\tt +j}=(1,0,0,0,-1,1,-1); {\tt Y}=(1,0,0,0,-1,-1,1); {\tt V}=(-1,0,0,0,1,1,1); {\tt v}=(1,0,0,1,0,1,1); {\tt h}=(1,0,0,1,0,1,-1); {\tt ++D}=(1,0,0,1,0,-1,1); {\tt +[}=(1,0,0,1,0,-1,-1); {\tt +S}=(1,0,0,1,1,0,1); {\tt +F}=(1,0,0,1,1,0,-1); {\tt +m}=(1,0,0,1,1,1,0); {\tt z}=(1,0,0,1,-1,0,1); {\tt +\#}=(1,0,0,1,-1,-1,0); {\tt N}=(1,0,0,-1,0,1,1); {\tt ++P}=(1,0,0,-1,0,1,-1); {\tt +e}=(1,0,0,-1,0,-1,1); {\tt i}=(-1,0,0,1,0,1,1); {\tt p}=(1,0,0,-1,1,0,1); {\tt B}=(1,0,0,-1,1,0,-1); {\tt ++J}=(1,0,0,-1,1,1,0); {\tt +K}=(1,0,0,-1,-1,0,1); {\tt +G}=(-1,0,0,1,1,0,1); {\tt @}=(-1,0,0,1,1,1,0); {\tt ++N}=(1,0,1,0,0,1,1); {\tt C}=(1,0,1,0,0,-1,1); {\tt +:}=(1,0,1,0,1,0,1); {\tt +X}=(1,0,1,0,1,0,-1); {\tt ++G}=(1,0,1,0,1,1,0); {\tt \textless }=(1,0,1,0,1,-1,0); {\tt j}=(1,0,1,0,-1,0,1); {\tt O}=(1,0,1,0,-1,0,-1); {\tt +q}=(1,0,1,0,-1,-1,0); {\tt ++5}=(1,0,1,1,0,0,1); {\tt ++1}=(1,0,1,1,0,0,-1); {\tt ++L}=(1,0,1,1,0,-1,0); {\tt +\textless }=(1,0,1,1,-1,0,0); {\tt )}=(1,0,1,-1,0,0,1); {\tt J}=(1,0,1,-1,0,0,-1); {\tt \{}=(1,0,1,-1,0,1,0); {\tt +v}=(1,-1,1,1,0,0,0); {\tt [}=(1,0,-1,0,0,1,1); {\tt +7}=(1,0,-1,0,0,1,-1); {\tt ++K}=(1,0,-1,0,0,-1,1); {\tt q}=(-1,0,1,0,0,1,1); {\tt k}=(1,0,-1,0,1,0,1); {\tt +w}=(1,0,-1,0,1,0,-1); {\tt +\textbackslash}=(1,0,-1,0,1,1,0); {\tt +\&}=(1,0,-1,0,1,-1,0); {\tt +n}=(1,0,-1,0,-1,0,1); {\tt +H}=(-1,0,1,0,1,0,1); {\tt =}=(-1,0,1,0,1,1,0); {\tt \%}=(1,0,-1,1,0,0,1); {\tt -}=(1,0,-1,1,0,0,-1); {\tt +\textgreater }=(1,-1,1,0,-1,0,0); {\tt ++7}=(1,0,-1,-1,0,0,1); {\tt +\}}=(-1,0,1,1,0,0,1); {\tt ++H}=(1,0,-1,-1,0,1,0); {\tt +?}=(-1,0,1,1,0,1,0); {\tt +8}=(1,-1,1,0,1,0,0); {\tt +O}=(1,-1,1,0,0,-1,0); {\tt +i}=(1,-1,1,0,0,1,0); {\tt +'}=(1,-1,1,0,0,0,-1); {\tt +x}=(1,1,0,0,0,1,1); {\tt ++O}=(1,1,0,0,1,0,-1); {\tt ++B}=(1,1,0,0,1,1,0); {\tt ++M}=(1,1,0,0,-1,0,1); {\tt ++Q}=(1,1,0,0,-1,1,0); {\tt +Y}=(1,1,0,0,-1,-1,0); {\tt +9}=(1,1,0,1,0,0,1); {\tt +$\vert$}=(1,1,0,1,0,1,0); {\tt K}=(1,1,0,1,0,-1,0); {\tt +E}=(1,1,0,1,1,0,0); {\tt +]}=(1,-1,1,0,0,0,1); {\tt *}=(1,1,0,1,-1,0,0); {\tt +A}=(-1,1,0,1,1,0,0); {\tt +o}=(1,1,0,-1,0,0,-1); {\tt +V}=(1,1,0,-1,0,1,0); {\tt w}=(1,1,0,-1,1,0,0); {\tt \&}=(1,-1,0,-1,1,0,0); {\tt +P}=(1,1,0,-1,-1,0,0); {\tt /}=(-1,1,0,1,0,1,0); {\tt +\_}=(1,1,1,0,0,0,1); {\tt '}=(1,1,1,0,0,1,0); {\tt +I}=(1,1,1,0,0,-1,0); {\tt \textbackslash}=(1,1,1,0,1,0,0); {\tt +\$}=(-1,1,0,1,0,0,1); {\tt +-}=(1,-1,0,-1,0,0,1); {\tt r}=(1,1,1,0,-1,0,0); {\tt ++3}=(1,1,1,1,0,0,0); {\tt ++R}=(1,-1,0,1,-1,0,0); {\tt +J}=(1,-1,0,1,1,0,0); {\tt +L}=(1,-1,0,1,0,-1,0); {\tt s}=(1,-1,0,1,0,1,0); {\tt ]}=(1,-1,0,1,0,0,-1); {\tt ++I}=(1,-1,0,1,0,0,1); {\tt +T}=(-1,1,0,0,1,1,0); {\tt !}=(1,1,1,-1,0,0,0); {\tt D}=(1,-1,0,0,-1,1,0); {\tt +`}=(-1,1,0,0,1,0,1); {\tt P}=(1,-1,0,0,1,-1,0); {\tt "}=(1,-1,0,0,1,1,0); {\tt +r}=(1,-1,0,0,0,1,-1); {\tt +B}=(1,1,-1,0,0,0,-1); {\tt ++2}=(1,1,-1,0,0,1,0); {\tt ++6}=(1,1,-1,0,1,0,0); {\tt $\vert$}=(1,1,-1,0,-1,0,0); {\tt E}=(1,1,-1,1,0,0,0); {\tt +s}=(1,1,-1,-1,0,0,0).

\subsection*{9-dim}

{\bf 47-16} {\tt 123PQSUkl, 12BCJLVWb, 13DEMOYab, 1BQRVXdei, 1CPRZafgi, 1DMNWXghj, 1EKLYZcdj, 1IJKTUfhl, 1INOSTcek, 23GHJLVWb, 45EFJKTUb, 46GIJKTUb, 56ABPQSUb, 78ACPQSUb, 79HINOSTb, 89DFQRVXb.} {\tt 1}=(0,0,0,0,0,0,0,0,1); {\tt 2}=(1,0,0,0,0,0,0,0,0); {\tt 3}=(0,0,1,0,0,0,0,0,0); {\tt 4}=(1,-1,1,0,0,0,0,0,1); {\tt 5}=(-1,1,1,0,0,0,0,0,1); {\tt 6}=(1,1,1,0,0,0,0,0,-1); {\tt 7}=(1,-1,1,0,0,0,0,0,-1); {\tt 8}=(1,1,-1,0,0,0,0,0,-1); {\tt 9}=(1,1,1,0,0,0,0,0,1); {\tt A}=(1,0,0,0,0,0,0,0,1); {\tt B}=(0,1,-1,0,0,0,0,0,0); {\tt C}=(0,1,1,0,0,0,0,0,0); {\tt D}=(1,-1,0,0,0,0,0,0,0); {\tt E}=(1,1,0,0,0,0,0,0,0); {\tt F}=(0,0,1,0,0,0,0,0,-1); {\tt G}=(0,1,0,0,0,0,0,0,1); {\tt H}=(0,1,0,0,0,0,0,0,-1); {\tt I}=(1,0,-1,0,0,0,0,0,0); {\tt J}=(0,0,0,1,1,0,0,0,0); {\tt K}=(0,0,0,1,-1,0,0,0,0); {\tt L}=(0,0,0,0,0,1,1,0,0); {\tt M}=(0,0,0,1,0,1,0,0,0); {\tt N}=(0,0,0,0,1,0,1,0,0); {\tt O}=(0,0,0,0,1,0,-1,0,0); {\tt P}=(0,0,0,0,1,-1,0,0,0); {\tt Q}=(0,0,0,0,1,1,0,0,0); {\tt R}=(0,0,0,1,0,0,1,0,0); {\tt S}=(0,0,0,1,0,0,0,0,0); {\tt T}=(0,0,0,0,0,1,0,0,0); {\tt U}=(0,0,0,0,0,0,1,0,0); {\tt V}=(0,0,0,1,-1,1,-1,0,0); {\tt W}=(0,0,0,1,-1,-1,1,0,0); {\tt X}=(0,0,0,1,1,-1,-1,0,0); {\tt Y}=(0,0,0,1,1,-1,1,0,0); {\tt Z}=(0,0,0,1,1,1,-1,0,0); {\tt a}=(0,0,0,-1,1,1,1,0,0); {\tt b}=(0,0,0,0,0,0,0,1,0); {\tt c}=(1,-1,1,0,0,0,0,-1,0); {\tt d}=(1,-1,-1,0,0,0,0,1,0); {\tt e}=(1,1,1,0,0,0,0,1,0); {\tt f}=(1,-1,1,0,0,0,0,1,0); {\tt g}=(1,1,-1,0,0,0,0,1,0); {\tt h}=(1,1,1,0,0,0,0,-1,0); {\tt i}=(1,0,0,0,0,0,0,-1,0); {\tt j}=(0,0,1,0,0,0,0,1,0); {\tt k}=(0,1,0,0,0,0,0,-1,0); {\tt l}=(0,1,0,0,0,0,0,1,0).

\medskip 

{\bf 19-5}\quad {\tt 678125493, 349CDFGHE, EFGHIJ786, ABCD56789, IJAB12789.}

\medskip 
{\bf 7(44)-6} \quad {\tt SU(CDEFOQ6), 1G42U(8H95), 1S(AIMSVXbi), 472(acefhK), G72(LWYZdg), 4U(BJ3NPRT).} {\tt 1}=(0,0,0,0,0,0,0,1,0); {\tt 2}=(0,0,0,0,0,0,1,0,0); {\tt 3}=(0,0,0,0,1,1,0,0,0); {\tt 4}=(0,0,0,1,0,0,0,0,1); {\tt 5}=(0,0,0,1,0,0,0,0,-1);  {\tt 6}=(0,0,0,1,0,0,-1,1,0); {\tt 7}=(0,0,1,0,0,0,0,1,0); {\tt 8}=(0,0,1,0,0,1,0,0,0); {\tt 9}=(0,0,1,0,0,-1,0,0,0); {\tt A}=(0,0,1,0,-1,0,1,0,0);  {\tt B}=(0,0,1,0,-1,1,1,0,0); {\tt C}=(0,0,-1,-1,1,1,0,1,1); {\tt D}=(0,0,1,-1,-1,-1,0,1,1); {\tt H}=(0,1,0,0,1,0,0,0,0); {\tt E}=(0,1,0,0,1,-1,1,1,-1);  {\tt F}=(0,1,0,0,1,-1,-1,-1,1); {\tt G}=(0,1,0,0,-1,0,0,0,0); {\tt I}=(0,1,0,1,1,1,1,0,1); {\tt J}=(0,1,0,-1,0,0,0,-1,1); {\tt K}=(1,-1,1,1,-1,1,0,-1,-1);  {\tt L}=(0,1,0,-1,1,1,0,0,0); {\tt M}=(0,1,0,-1,1,-1,1,0,-1); {\tt N}=(0,1,-1,0,0,0,1,1,0); {\tt O}=(0,1,1,1,0,1,1,0,1); {\tt P}=(0,1,1,1,1,-1,1,-1,-1);  {\tt Q}=(0,1,1,-1,0,1,-1,0,-1); {\tt R}=(0,1,-1,0,0,0,1,1,0); {\tt S}=(0,-1,1,0,1,0,0,0,0); {\tt U}=(1,0,0,0,0,0,0,0,0); {\tt V}=(1,0,0,0,0,-1,0,0,1);  {\tt W}=(1,0,0,1,0,1,0,0,1); {\tt X}=(1,0,0,-1,0,1,0,0,0); {\tt Y}=(1,0,0,-1,0,-1,0,0,1); {\tt Z}=(1,0,1,0,0,0,0,-1,-1); {\tt a}=(1,1,0,0,-1,-1,0,0,0); {\tt b}=(1,1,1,1,0,0,-1,0,-1); {\tt c}=(1,1,1,-1,1,1,0,-1,1); {\tt d}=(-1,1,1,1,1,-1,0,-1,1); {\tt e}=(1,-1,-1,-1,-1,1,0,1,1); {\tt f}=(1,1,-1,1,1,1,0,1,-1); {\tt g}=(1,1,-1,1,1,-1,0,1,-1); {\tt h}=(1,-1,0,0,1,-1,0,0,0); {\tt i}=(1,-1,-1,1,0,0,1,0,-1).}

\bigskip

MMPHs in Fig.~4(b,c) ({\bf 44-6} and {\bf 7-6} above) of the main body
of the paper illustrate why all vertices are needed for a consistent
implementation of MMPHs. Let us assign `1' to {\tt S} and {\tt 2} in
Fig.~4(c), i.e.~to {\bf 7-6}. The hyperedge {\tt 4U} then provides us
with the contextual contradiction since `0' is assigned to both vertices
in it. But the hyperedge {\tt 4U} would not be a hyperedge without
all the other vertices ({\tt B,J,3,N,P,R,T}) necessary for preparing
and measuring it, since without them it would be just the
orthogonality between {\tt 4} and {\tt U} and would belong to the
hyperedge {\tt 1G42U}, i.e., would not be a separate hyperedge. It
is also indicative that Fig.~4(c), i.e.~{\bf 7-6}, is critical with any
or all of {\tt a,c,e,f,h,K} considered in hyperedge {\tt 472} while it
ceases to be contextual as soon as we consider any of
{\tt B,J,3,N,P,R,T} in {\tt 4U} since we can then assign a 3rd `1'
to one of them.

\end{widetext}


\begin{widetext}

\section*{Appendix 1:  {\em Physical Review Letter}
  Editor's 1st decision, 1st referee report, and
  my response to it}

[Appendix 3 below contains Physical Review Letter
  Editor's 2nd decision, 2nd referee report, and
  my response to it; Appendices 4 and 5 at the end contains
  Physical Review Letter Editor's 3rd decision and my
  comment on it.] 

  \subsubsection*{\rm Mladen Pavi\v ci\'c}
  
  \subsection*{Editor's 1st decision}

  Journal:\quad Physical Review Letters; \quad
  Paper:\quad LP17301; \quad Received:\quad 17Feb2022

  16Jun22 \phantom{16Jun22} Editorial decision

  19May22 16Jun22 Review editor concludes response unlikely

  08Mar22 19May22 Review request to referee; report received

  21Apr22 22Apr22 Review request to referee; message received
  (not a report)

  29Mar22	30Mar22	Review request to referee; message
  received (not a report)

  08Mar22	09Mar22	Review request to referee; message
  received (not a report)

Date: Thu, 16 Jun 2022 14:00:15 +0000

Re: LP17301
    Automated generation of arbitrarily many Kochen-Specker and other
    contextual sets in odd dimensional Hilbert spaces
    by Mladen Pavi\v{c}i\'{c} and Norman D. Megill

Dear Prof. Pavicic,

This manuscript has been reviewed by our referee. A critique from the
report appears below. Based on this we judge that the work probably
warrants publication in some form, but does not meet the Physical
Review Letters criteria of impact, innovation, and interest.

The paper, with revision as appropriate, might be suitable for
publication in one of the topical Physical Review journals (e.g.,
Physical Review A) or Physical Review Research. The editors of that
journal will make the decision on publication, and may seek further
review. However, our complete file is available.

Yours sincerely,

Robert Garisto (he/him/his)
Editor
Physical Review Letters

P.S. Another referee we consulted was not able to review your
manuscript.

\bigskip

---------------------------
{\bf Report of the 1st Referee -- LP17301/Pavicic}
---------------------------

\medskip

*OVERALL EVALUATION*

\smallskip
This article aims to present new methods to generate large numbers
of examples of Kochen-Specker sets in odd-dimensional Hilbert spaces.
I found the manuscript lacking in a number of aspects, which I now
highlight. (In the ``Detailed Comments'' section below, I go into
more detail on some of these issues, illustrating them with concrete
passages, among other minor comments.)

1. The manuscript does not provide enough motivation nor does it try
to make a case for the relevance of this work and its appeal to a
broad audience such as PRL's. It claims that "development of quantum
computation and communication, recently shown to be supported by
contextuality, arguably asks for an abundant supply of contextual
sets", but the authors do not bother to explain why such sets might be
necessary. I'm not claiming that there isn't a case to be made. There
likely is, but the point is that the manuscript does this very weakly.

2. The presentation falls short of the standard expected in a research
article, especially in a journal as good as PRL, in terms of clarity
and rigor. Various passages are written in a confusing fashion. For
example, it is often unclear from the text that a passage is to be
taken as the definition of a new concept (e.g. when the notion of
"coordinatization" is introduced). On the other hand, e.g. in the
definition of MMPH hypergraph, some extra commentary is written as
part of the definition when it does not belong there, namely item (v).
Overall, the presentation is often imprecise (see detailed comments
below for more examples).

3. The choice of material included in the text is odd, and it is
inadequate for a journal article. At points, the authors choose to
give too much information on irrelevant implementation details (e.g.,
the internal ASCII encoding of the graphs being manipulated, or the
name of functions in their computer code). In contrast, and more
worryingly, the text is quite vague in the description of the actual
methods/algorithms being implemented (especially M3 which seems to be
the less obvious of them). This is problematic for two reasons. First,
it detracts from the interest that the manuscript may have to a
journal's audience: most readers would read the paper to learn the
main ideas underlying the methods so they could replicate and adapt
them (anyone keen on learning implementation details would refer to
the code's documentation and/or to the code itself). Moreover, this
vagueness makes it hard for referees to judge the authors'
contributions in terms of correctness, novelty, etc.

For the above reasons, I am unfortunately not able to recommend this
article for publication in PRL. I believe that the results may be of
interest to a more specialized audience. I would therefore encourage
the authors to submit a carefully *revised* version, with a more
detailed presentation of the main novel technical contributions, to a
journal such as PRA.

\medskip
*DETAILED COMMENTS*

\smallskip
(some more important than others, in order of appearance in the
manuscript)

– Abstract, line 1-2:

"Development of quantum computation and communication, recently shown
to be supported by contextuality, *arguably asks for an abundant
supply of contextual sets*"

Contextuality has been shown to underlie quantum advantage in various
computation and communication tasks and protocols. However, the second
part of the sentence does not follow directly from this. That is, why
does the development of quantum computation and communication ask for
an *abundant supply of contextual sets*? This may well be the case,
but (1) it is not an obvious consequence of the results mentioned from
the literature, and (2) the authors fail to argue for this. Given that
this is presented as the main motivation for this work and for its
broad appeal, I would expect to see a better case made for this.

– Page 1, col. 1, l. 7:

"Qubits (quantum bits) are two-dimensional qudits (d = 2) and their
tensor products build the corresponding even-dimensional Hilbert
spaces."

While the Hilbert space corresponding to an n-qubit system (i.e., an
n-fold tensor of $C^2$) is even dimensional, not all even-dimensional
Hilbert spaces are of this form. The Hilbert space of multiple qubits
has dimension a power of two. One does not get, for example, a
6-dimensional Hilbert space in this fashion. Some of the discussion
around even vs odd dimension in the article seems to misleadingly
conflate even dimensionality with being realizable with qubits.

– Page 1, col. 1, l. 21:

"and *several* in 5- to 11-dim spaces"

This is a rather vague statement: what is meant by "several"?

– Page 1, col. 1, l. 22:

"[...]of Kochen-Specker contextual sets only four [...] were
explicitly provided for particular sets"

It is unclear what the authors mean by "were explicitly provided for
particular sets"?

– Page 1, col. 1, l. 38:

"An MMPH is a connected hypergraph k-l with k vertices and l
hyperedges in which [...] every hyperedge contains at least 2 and *at
most n* vertices"

The number "n" here appears out of nowhere. Only "k" and "l" are bound
by the definition of "hypergraph k-l", but "n" means nothing in this
context and therefore the requirement that every hyperedge contain at
most n vertices seems to be an empty one.

– Page 1, col.2, l. 1:

"(v) graphically, vertices are represented as dots and hyperedges as
(curved) lines passing through them."

This is given as one of the conditions in the definition of MMPH
graphs, but in reality it is not a condition on hypergraphs. It is
just an explanation of the convention used to draw hypergraphs to ease
the readability of Figure 2, for example. It should be separated from
the definition of MMPH hypergraph.

– Page 1, col. 2, l. 4-10:

"We encode MMPHs by means of the following 90 ASCII characters: [list
of ASCII characters] [...]"

This is the kind of implementation detail that is completely
irrelevant to the potential readers of this paper. Its inclusion is a
waste of space that could be used to convey much more interesting
information about the methods implemented. This does not concern any
substantial/significant algorithmic choice, but simply a matter of
convention chosen for the internal representation. It's the kind of
detail that is typically hidden from the user even in the
documentation of a well-written library, and it's even less relevant
in the context of a journal article, which ought to focus on
explaining the methods instead.

– Page 1, col. 2, l. 14-16:

"An MMPH is a special kind of a general hypergraph in the sens that
none of the aforementioned points (i-v) holds for it."

The meaning of this sentence is unclear. What is the referent of the
pronoun "it" at the end of the sentence? (By the way, there is also a
minor typo: "sens" should be spelt "sense".)

– Page 1, col. 2, l. 23-25:

"Orthogonality between vertices in an MMPH space just means that the
vertices are contained in their hyperedges".

This is yet another sentence whose meaning isn't clear. What is meant
by "the vertices are contained in *their* hyperedges"? In fact, this
whole paragraph about "coordinatization" is written in a confusing
fashion. It seems that the intention is for it to be a definition of
MMPH with coordinatization, but the definition proper appears as
seemingly an afterthought, in the clause starting with "i.e.".

– Page 1, col. 2, l. 34 (and after):

The acronym "MMPH" is long and awkward enough as it is. The strings
"NBMMPH" and "(N)BMMPH", and later even "KS NBMMPH", are beyond what
is reasonable to expect a reader to parse. It would be much better to
keep the full phrase "non-binary". I would even go further to suggest
finding an alternative to MMPH. Also because two of the initials in
this acronym refer to the authors of the present paper...

– Page 2, col. 1, l. 2-6:

"NBMMPHs are nonclassical since they do not allow assignments of
predefined 0s and 1s to their vertices; therefore, we consider them to
be contextual. BMMPHs are classical since they do allow such an
assignment; therefore, we consider them to be noncontextual."

This sentence reads awkwardly. What is meant by "*we consider* them to
be (non)contextual". Is this supposed to be a definition?

– Page 2, col. 1, l. 11–12:

"When conditional or unconditional contextual operators [...]"

The terms conditionally contextual, unconditionally contextual, and
conditionally noncontextual are never defined or even explained
informally.

– Page 2, col. 1, l. 23-43:

"To generate (N)BMMPHs in the odd-dim spaces we make use of three
methods [...]"

These short paragraphs are what is included as a description of the
methods used. The descriptions of each of M1 to M3 provided are
extremely brief and lacking in detail. A more detailed exposition of
the methods and some examples illustrating them, especially for M3,
would be useful and make the paper more interesting to potential
readers. Also, it would be good to see more details on how the code
deals with / finds / manipulates coordinatizations (i.e., realizations
with quantum states and measurements) for the graphs in question.

–Page 2, col. 1, l. 57 - col. 2, l. 1:

"using our program *MMPSstrip*" [...] "stripping them to the critical
KS MMPHs by *States01*"

The name of the program or routine that performs a given task (such as
"MMPSstrip" or "States01") is totally irrelevant here. On the other
hand, it would be more interesting to convey what these programs are
actually doing: how are the edges added? randomly?, how are they
stripped to critical sets?, etc.

– Page 2, col. 2, Fig. 1:

This figure is hard to interpret. The caption does not adequately
explain, e.g., what do the axes represent, or what are the shaded
black stripes. The same applies to Fig 3 and Fig 4 later on.

– Page 2, col. 2, l. 13-14:

"When applying M3 we obtain that Bub’s is the only 49-36 NBMMPH and
that there are no smaller KS ones for the considered vector
components."

Again, this statement is unnecessarily vague. What is meant by "for
the considered vector components"? Which vector components are being
considered which allow the authors to make this statement? This goes
back, I believe, to the overly vague explanation given of M3.

– Page 5, col.1, l. 26-30

"The methods are especially needed in odd dimensional Hilbert spaces
since, in contrast to even dimensional ones, we cannot make use of
polytopes, Pauli operators, qubit states, parities, and other
approaches specific to qubit spaces".

Are analogous methods not available for qudits? Is there a reason why
this is the case?

\bigskip
---------------------
{\bf Author's response to the Report of the 1st Referee -- LP17301/Pavicic}
---------------------

\medskip

The referee is not at home with the theory of hypergraphs on which
our article is based; none of her/his comments holds water.
So, an acceptance of her/his report as the only ground for a
rejection of the present article would likely be detrimental for
the reputation of the {\em Physical Review Letters\/} in this field.

In the revised version of the article ( available at
https://arxiv.org/abs/2202.08197 ) I made some changes that
clarify the points the referee misinterpreted, misunderstood,
or twisted. In my response below, I indicate such changes by
``*****************''. Referee's comments I indicate by ``$>$''.

\medskip

$>$ *OVERALL EVALUATION*

\smallskip
$>$ This article aims to present new methods to generate large
numbers of examples of Kochen-Specker sets in

$>$ odd-dimensional Hilbert spaces.

\smallskip

Definitely not. In the summary we clearly stated our aim:
``Our goal is \dots to establish general methods for automated
generation of NBMMPHs [of which the Kochen-Specker sets are the
subsets] in any dimension for any possible future application and
implementation, e.g., in quantum computation and communication.'' 
In the present article we focus on the odd-dimensional spaces
since we obtained analogous results for the even-dimensional
spaces in the previous publications of ours. Also, in the
even-dimensional spaces there are numerous other methods by
which one can achieve automated generation of NBMMPHs, while in
the odd-dimensional spaces apparently only the ones presented in
this article are available.

\smallskip 
$>$ 1. The manuscript does not provide enough motivation nor
does it try to make a case for the relevance of this 

$>$ work and its appeal to a broad audience such as PRL's.
It claims that ``development of quantum computation

$>$ and communication, recently shown to be supported
by contextuality, arguably asks for an abundant supply of

$>$ contextual sets,'' but the authors do not bother to explain
why such sets might be necessary. I'm not claiming 

$>$  that there isn't a case to be made. There likely is, but
the point is that the manuscript does this very weakly.

\smallskip

On the contrary, the motivation is clearly expressed in the second
paragraph of the paper: ``none of the methods [previously
employed in the even-dimensional spaces] is available in the
odd-dimensional spaces. In this paper, we offer several methods
for automated generation of arbitrarily many KS sets as well as
contextual non-KS ones in $n=3,5,7,9$ dimensional spaces.''

The sentence: ``quantum computation and communication \dots
arguably asks for an abundant supply of contextual sets''
is taken from the abstract which is limited in size. In the
very body of the article, it is substantiated by the
motivation sentences cited above. 

**** In the present version of the paper I expanded the Abstract 
so as to make the first and the last sentences read: ****

**************  Development of quantum computation and communication,
recently shown to be supported by contextuality, arguably asks for a
requisite supply of contextual sets. ************** 

************** In this paper, we offer three methods for
automated generation of arbitrarily many contextual KS and
non-KS sets in any dimension for possible future application
and implementation, e.g., in quantum computation and
communication, which can be applied in both even and odd
dimensional spaces and we employ them to obtain millions of
KS and other contextual sets in dimensions 3, 5, 7, and 9.
**************

\smallskip
$>$ 2. The presentation falls short of the standard expected in
a research article, especially in a journal as good as

$>$ PRL, in terms of clarity and rigor. Various passages are
written in a confusing fashion.

\smallskip
After so many articles we published in {\em Phys.~Rev.~Lett.,
  Scientific Reports, Phys.~Rev.~A $\&$ D, Optics Express,
  J.~Math.~Phys., J.~Phys.~A, J.~Opt.~Soc.~Am.~B, Entropy,
  Phys.~Lett.~A, Found.~Phys., Ann.~H.~Poincare \dots\/} over
almost half a century, enters a referee to announce that we do
not know how to write an article. The article is written up to
the standards of the PRL and nothing is written in a confusing
fashion, only the referee does not grasp the meaning of the
presented material since she/he is not at home with the
theory of hypergraphs. 

\smallskip 
$>$ For example, it is often unclear from the text that a
passage is to be taken as the definition of a new concept

$>$ (e.g.~when the notion of ``coordinatization'' is introduced).

\smallskip 
{\em Wiktionary\/} has a single item for: 
{\em coordinatization\/} (plural {\em coordinatizations\/}):
(mathematics) ``{\em The application of coordinates to a system\/},''
and in the 3rd paragraph of the right column on p.~1 we
additionally clarify: ``The MMPHs above are defined without a
coordinatization, i,e., neither their vertices nor their
hyperedges are related to either vectors or operators.'' So, no
``new concept'' is introduced. The ``coordinatization'' is a word
from a dictionary the referee should have consulted. 

\smallskip 
$>$ On the other hand, e.g. in the definition of MMPH hypergraph,
some extra commentary is written as a part of 

$>$ the definition when it does not belong there, namely item (v).

\smallskip
The referee is obviously not at home with the theory of
hypergraphs. E.g., Berge,
  {\em Hypergraphs: Combinatorics of Finite Sets\/},
  North-Holland (1989); Ch.~1, Sec.~1, p.~1, 4th paragraph:
  ``A hypergraph may be drawn as a set of points representing the
  vertices [and a hyper]edge is represented \dots by a single closed
  curve enclosing the vertices''---as a part of his definition.
  There is no particular kind of a ``definition'' of ours which
  would differ from Berge's kind except that we introduce,
  in point (v), a curve passing through vertices instead of
  Berge's curve which encloses them. 

\smallskip
$>$ Overall, the presentation is often imprecise (see detailed
comments below for more examples).

\smallskip
The referee fails to substantiate her/his claim in any of
her/his comments. 

\smallskip
$>$ 3. The choice of material included in the text is odd, and
it is inadequate for a journal article. At points, the 

$>$ authors choose to give too much information on irrelevant
implementation details (e.g., the internal ASCII 

$>$ encoding of the graphs being manipulated, or the name of
functions in their computer code).

\smallskip
Again, the referee clearly shows that she/he is not at home with
the theory of hypergraphs. Berge, Ch.~1, Sec.~1, p.1, 5th
paragraph: ``One may also define a hypergraph by its
incidence matrix, with columns representing the [hyper]edges and
rows representing the vertices (cf. Figure 1, p.~2).''
We, instead of an incidence matrix, make use of MMPH ASCII strings.
Some kind of codification is unavoidable and our MMPH string
codification is an integral part of the MMPH language and these
strings are generated by our algorithms and programs and they
then serve us as further direct inputs to our algorithms and
programs whose outputs are all MMPH properties and features including
graphical representations of the obtained MMPHs. The difference
between incidence matrices and MMPHs is obvious; for instance, the
MMPH string of Fig.~2(c) can be written in
a single line, while the incidence matrix takes 37 columns and
51 lines. Also an MMPH string can be put in any of our programs
while no such programs for arbitrary incidence matrices are known
to us. And then enters the referee to proclaim the MMPH
codification ``irrelevant.'' 

\smallskip
**** In the present version of the paper I add the following
sentences to the right column, p.~1: ****

************** So encoded single ASCII characters (possibly
prefixed by {\tt +}'s) represent vertices; e.g.,
{\tt 1}, or {\tt q}, or {\tt +++A}. Put one after another (without
spaces) they represent hyperedges; e.g., {\tt 123}, or {\tt CZk1@=a},
or {\tt 1+1+++1Dd}. In a representation of an MMPH, hyperedges are
organized in a string in which they separated by commas; each string
ends with a period; e.g., the string
{\tt 123,345,567,789,9A1.}~represents a noncontextual 3-dim MMPH
pentagon. **************

************** We stress here that MMPH strings are generated by our
algorithms and programs and then processed to yield all properties
and features of MMPHs, including their graphical representations.
 ************** 

\smallskip
$>$ In contrast, and more worryingly, the text is quite vague in the
description of the actual methods/algorithms

$>$ being implemented (especially M3 which seems to be
the less obvious of them).

\smallskip
I'm flabbergasted. What can be simpler than, p.~2:
``{\bf M3}  consists in combining simple vector components
so as to exhaust all possible collections of $n$ mutually
orthogonal $n$-dim vectors.'' For instance 3-dim components
$(0,1)$ yield: {\tt 123.} {\tt 1}=(0,0,1), {\tt 2}=(0,1,0),
{\tt 3}=(1,0,0); 4-dim $(0,\pm1)$ yield Peres' set 24-24;
etc.

\smallskip
$>$ This is problematic for two reasons. First, it detracts from
the interest that the manuscript may have to a

$>$ journal's audience: most readers would read the paper to learn
the main ideas underlying the methods so they

$>$ could replicate and adapt them (anyone keen on learning
implementation details would refer to the code's

$>$ documentation and/or to the code itself). Moreover, this
vagueness makes it hard for referees to judge the

$>$  authors' contributions in terms of correctness, novelty, etc.

\smallskip
What is problematic here are the motives for such twists on the
part of the referee. The first sentence seems to contradict
itself: ``most readers would read the paper to learn the main
ideas underlying the methods so they could replicate and adapt
them,'' but they would not like ``to learn the code itself.''
Note that without the MMPH language the results obtained in the
article are inaccessible as they were in the last half of
century.

In the second sentence, ``this vagueness''---read MMPH
language---makes it hard for [malevolent] referees to judge the
authors' contributions in terms of correctness, novelty, etc.
As if it is not transparent that the correctness and novelty of
all results can be verified directly at the results presented
in the article and its appendices or at our repository
http://puh.srce.hr/s/Qegixzz2BdjYwFL where all algorithms and
programs are freely available. 

\smallskip
$>$ For the above reasons, I am unfortunately not able to
recommend this article for publication in PRL.

$>$ I believe that the results may be of interest to a more
specialized audience. 

\smallskip
The researchers were striving to obtain these results for at
least quarter of century and it is regretful that the editor
bought such a recommendation.

\smallskip
$>$ I would therefore encourage the authors to submit a
carefully *revised* version, with a more detailed 

$>$ presentation of the main novel technical contributions,
to a journal such as PRA.

\smallskip
Our algorithms do require sophisticated programs but that does not
make the contributions ``technical,'' because there is no other way
to obtain them.

\medskip
$>$ *DETAILED COMMENTS*

\smallskip
$>$ -- Abstract, line 1-2:

$>$ "Development of quantum computation and communication, recently
shown to be supported by contextuality,

$>$ *arguably asks for an abundant supply of contextual sets*"
Contextuality has been shown to underlie quantum

$>$ advantage in various computation and communication tasks 
and protocols. However, the second part of the

$>$ sentence does not follow directly from this. That is, why does
the development of quantum computation and

$>$ communication ask for an *abundant supply of contextual sets*?
This may well be the case, but (1) it is not an

$>$ obvious consequence of the results mentioned from the
literature, and (2) the authors fail to argue for this. Given

$>$ that this is presented as the main motivation for this work
and for its broad appeal, I would expect to see a

$>$ better case made for this.

\smallskip
Common scientific sense tells us that when ``what gives quantum
computers that extra oomph over their classical digital counterparts
\dots is an intrinsic, measurable aspect of quantum
mechanics called contextuality'' [1,2]  we should have more than
a few sets per dimension. Those recent result is so well-known that we
considered it far outside the scope of the present paper to elaborate
on it or to even add references.

\smallskip
**** Still in the present version of the paper I added references
[1,2,3,4] and I substituted ``requisite'' for ``abundant'' ****

\smallskip
$>$ -- Page 1, col. 1, l. 7:

\smallskip
$>$ "Qubits (quantum bits) are two-dimensional qudits ($d = 2$) and
their tensor products build the corresponding

$>$ even-dimensional Hilbert spaces." While the Hilbert space
corresponding to an n-qubit system (i.e., an $n$-fold

$>$ tensor of $C^2$) is even dimensional, not all even-dimensional
Hilbert spaces are of this form. The Hilbert space

$>$ of multiple qubits has dimension a power of two. One does not
get, for example, a 6-dimensional Hilbert space

$>$ in this fashion. Some of the discussion around even vs odd
dimension in the article seems to misleadingly conflate

$>$ even dimensionality with being realizable with qubits.

\smallskip
This is ridiculous. We first write: ``[Qubit] tensor product
build the corresponding even-dimensional Hilbert spaces.''
Every undergraduate student understands that that means:
$2^n$, $n=2,3,4,\dots$. Then we write: ``in [5–9] billions of
contextual sets in 4-, 6-, 8-, 16-, and 32-dim Hilbert spaces,
predominantly related to qubits, were generated.'' And again
every undergraduate student would immediately recognize that
``predominantly'' refers to 4, 8, 16, and 32, but not to 6.
It seems to be a failed weaponized pedantry ? 

\smallskip
$>$ -- Page 1, col. 1, l. 21:

$>$ ``and *several* in 5- to 11-dim spaces'' This is a rather
vague statement: what is meant by ``several''?

\smallskip
Again, ridiculous. By {\em Webster Dictionary\/} ``several''
means ``more than two but fewer than many.''

\smallskip
$>$ -- Page 1, col. 1, l. 22:

$>$ ``[...] of Kochen-Specker contextual sets only four [...] were
explicitly provided for particular sets''

$>$ It is unclear what the authors mean by ``were explicitly
provided for particular sets?''

\smallskip
Is the referee serious? The full sentence reads: ``of Kochen-Specker
contextual sets only four in the 3-dim space [8,10] and several in
5- to 11-dim spaces [11–13] were explicitly provided for particular
sets,'' meaning that in those references sets with explicitly
given vectors were provided. I gave the sentence to a group of
students and no one had any problem with understanding the
meaning of ``explicitly,'' so I presume the reader would not
have it either. 

\smallskip
$>$ -- Page 1, col. 1, l. 38:

$>$ ``An MMPH is a connected hypergraph k-l with k vertices and l
hyperedges in which [...] every hyperedge

$>$ contains at least 2 and *at most n* vertices.'' The number "n"
here appears out of nowhere. Only "k" and "l"

$>$ are bound by the definition of "hypergraph k-l", but "n" means
nothing in this context and therefore the

$>$ requirement that every hyperedge contain at most n vertices
seems to be an empty one.

\smallskip
The referee again shows that she/he is not at home with the theory
of hypergraphs. The $n$ does not ``appear out of nowhere,'' but out
of the one but preceding sentence: ``In this paper, we offer
several methods for automated generation of arbitrarily many KS
sets as well as contextual non-KS ones in $n=3,5,7,9$" dimensional
spaces.''

\smallskip
**** Still in the present version of the paper I added ``$n$'' to
the cited sentence so as to now read: ****

************** An MMPH is a connected $n$-dim hypergraph $k$-$l$
with $k$ vertices and $l$ hyperedges \dots\ ************** 

\smallskip
$>$ -- Page 1, col.2, l. 1:

$>$ "(v) graphically, vertices are represented as dots and
hyperedges as (curved) lines passing through them."

$>$  This is given as one of the conditions in the definition
of MMPH graphs, but in reality it is not a condition on

$>$  hypergraphs. It is just an explanation of the convention
used to draw hypergraphs to ease the readability of

$>$  Figure 2, for example. It should be
separated from the definition of MMPH hypergraph.

\smallskip
Nowhere do we speak about a ``condition.'' It is just the point
within a definition in the same way in which similar points are
used in many hypergraph textbooks. We already clarified that
point (v) above when we compare it with a similar point used by
Berge, Ch.~1, Sec.~1, p.~1, 4th paragraph.

\smallskip
$>$ -- Page 1, col. 2, l. 4-10:

$>$ "We encode MMPHs by means of the following 90 ASCII characters:
[list of ASCII characters] [...]"

$>$ This is the kind of implementation detail that is completely
irrelevant to the potential readers of this paper.

$>$ Its inclusion is a waste of space that could be used to convey
much more interesting information about the

$>$ methods implemented. This does not concern any
substantial/significant algorithmic choice, but simply a

$>$ matter of convention chosen for the internal representation.
It's the kind of detail that is typically hidden

$>$ from the user even in the documentation of a well-written
library, and it's even less relevant in the context

$>$ of a journal article, which ought to focus on
explaining the methods instead.

\smallskip
Once again, the referee clearly shows that she/he is not at home
with the theory of hypergraphs at all. ``ASCII ``completely
irrelevant,'' ``waste of space,'' ``matter of convention,''
``a journal article ought to focus on explaining the methods
instead''\dots'' There are no methods which could generate the
contextual sets in odd dimensional spaces outside the MMPH
language based on ASCII codification either in our minds or
anywhere in the literature. MMPH language IS the method.
A ``{\em method instead\/}'' does not exist. As the title of
the paper stresses, the method consists in {\em automated\/}
computer-program-enabled generation of such contextual sets
based on the MMPH language. 

\smallskip

$>$ -- Page 1, col. 2, l. 14-16:

$>$  "An MMPH is a special kind of a general hypergraph in the
sense that none of the aforementioned points (i-v)

$>$ holds for it." The meaning of this sentence is unclear. What
is the referent of the pronoun "it" at the end of the

$>$ sentence?

\smallskip
``it'' = a general hypergraph. It is clear because points
(i-v) obviously hold for MMPH. 

\smallskip
$>$ -- Page 1, col. 2, l. 23-25:

$>$ "Orthogonality between vertices in an MMPH space just means
that the vertices are contained in their

$>$ hyperedges". This is yet another sentence whose meaning isn't
clear. What is meant by "the vertices are

$>$ contained in *their* hyperedges"?

\smallskip
For anyone who reads the paper so as to find out how MMPHs and
the orthogonality within them are structured it is clear what
is meant by *their*. Explanation is crystal clear: Let ABC be
a hyperedge; we can say, for vertices A,B,C, the hyperedge is
*their* hyperedge; so, the vertices A,B,C are mutually
orthogonal just by being contained in *their* hyperedge.
I'm convinced the reader will have much greater imagination
and insight than the present referee. 

\smallskip
$>$ In fact, this whole paragraph about "coordinatization" is
written in a confusing fashion. It seems that the

$>$ intention is for it to be a definition of MMPH with
coordinatization, but the definition proper appears as 

$>$ seemingly an afterthought, in the clause starting with "i.e.".

\smallskip
Confusion is seemingly with the referee. Our paragraph clearly
states that MMPHs are defined without any specification what
their vertices might be. When vectors, states, projectors, or
operators are assigned to vertices, then we speak about MMPHs
with coordinatization. The former MMPHs with bare unspecified
vertices are consequently MMPHs without coordinatization. 

\smallskip

$>$ -- Page 1, col. 2, l. 34 (and after):

$>$ The acronym "MMPH" is long and awkward enough as it is. The
strings "NBMMPH" and "(N)BMMPH", and 

$>$ later even "KS NBMMPH", are beyond what is reasonable to
expect a reader to parse. It would be much better 

$>$ to keep the full phrase "non-binary".

\smallskip
The term ``non-binary MMPH'' introduced and abbreviated to
NBMMPH on p.~2, is specific for this short article and the reader
will want to go back to its definition a couple of times while
reading it. So, ``non-binary MMPH'' vs.~NBMMP would not
make a difference in this respect; the full phrase would
only make the article longer. 

\smallskip
$>$ I would even go further to suggest finding an alternative to
MMPH. Also because two of the initials in this 

$>$ acronym refer to the authors of the present paper...

\smallskip
Is the referee serious? The acronym appears in more than 20 papers
since 2005 and no one of the former referees has ever brought it
into question. Besides, one of the authors (N.M.) is unfortunately
not alive any more. 

\smallskip

$>$ -- Page 2, col. 1, l. 2-6:

$>$ "NBMMPHs are nonclassical since they do not allow assignments of
predefined 0s and 1s to their vertices;

$>$ therefore, we consider them to be contextual. BMMPHs are
classical since they do allow such an assignment; 

$>$ therefore, we consider them to be noncontextual."
This sentence reads awkwardly. What is meant by 

$>$  "*we consider* them to be (non)contextual". Is this supposed to
be a definition?

\smallskip
What is that about? We simply state that we *consider* them in
the same way as they are *considered* to be of such a kind in the
literature over and over. For instance, ``In {\bf noncontextual}
hidden-variable theories the predetermined results of an observable
are independent of any other observables that are measured jointly
with it''\break [C. Brukner and M. Zukowski, Bell's Inequalities:
Foundations and Quantum Communication, in Handbook of Natural
Computing, Eds: G. Rozenberg, T.H.W. Baeck, and J.N. Kok, Springer
(2010), pp.~1413-1450]. {\bf Contextual} sets as ones that
do not admit predetermined values and {\bf noncontextual} as ones
that do go back to the original Kochen and Specker paper [20].

\smallskip
$>$ -- Page 2, col. 1, l. 11–12:

$>$ "When conditional or unconditional contextual operators [...]"
The terms conditionally contextual, 

$>$ unconditionally contextual, and conditionally noncontextual are
never defined or even explained informally.

\smallskip
Yes, the terms might have required some comments.

**** Instead, in the present version of the paper I substituted the
following sentence for the sentences containing the disputed terms
(3rd paragraph of the left column, on p.~2): ****

**************  When either state-dependent or state-independent
tests of operators defined on vertices of an NBMMPH with
$\kappa(i)\le n$ confirm the contextuality,
e.g.~\cite{yu-oh-12,cabello-bengtsson-12,cabello-svozil-18},
then the NBMMPH turns out to be contextual in all considered
cases so far. **************

State-independent contextuality (SIC) is a well-known term which
does not require any further explanation. 

\smallskip
$>$ -- Page 2, col. 1, l. 23-43:

$>$ "To generate (N)BMMPHs in the odd-dim spaces we make use of
three methods [...]" These short paragraphs 

$>$ are what is included as a description of the methods used.
The descriptions of each of M1 to M3 provided are

$>$ extremely brief and lacking in detail.

\smallskip
They are brief but not lacking in detail. {\bf M1}: dropping of
$m=1$ vertices means automated dropping of $m=1$; there is
nothing more to it. {\bf M2}: automated random addition of
hyperedges means automated random addition of hyperedges;
there is nothing more to it; {\bf M3}: combining simple vector
components so as to exhaust all possible collections of $n$
mutually orthogonal $n$-dim vectors is just that; e.g., (0,1)
components in a 3-dim space yield vectors: A=(0,0,1),
B=(0,1,0), C=(1,0,0) which form the hyperedge ABC; there is
nothing more to it. The referee apparently cannot believe that
everything turns out to be so simple. 
 
\smallskip
$>$ A more detailed exposition of the methods and some examples
illustrating them, especially for M3, would be

$>$ useful and make the paper more interesting to potential
readers. Also, it would be good to see more details on

$>$ how the code deals with / finds / manipulates
coordinatizations (i.e., realizations with quantum states and 

$>$ measurements) for the graphs in question.

\medskip
On the contrary. As I just stressed, what is useful and
interesting to the reader is to see how simple and straightforward
the algorithms are. Detailed examples of the three methods in all
considered dimensions are given in the body of the article.

**** As for the realizations/implementations, in the present version
of the paper I add the following sentences at the end of the
2nd paragraph of the left column, on p.~2): ****

************** The most direct implementation of an MMPH can be
carried out by measuring its vectors/states coming out of the gates
determined by its hyperedges. Operator based implementation,
e.g.~the one in which projectors determined by states/vertices
are projected on many other chosen states, is more demanding.
**************

\smallskip
$>$ -- Page 2, col. 1, l. 57 - col. 2, l. 1:

$>$ "using our program *MMPSstrip*" [...] "stripping them to the
critical KS MMPHs by *States01*." The name

$>$ of the program or routine that performs a given task (such as
"MMPSstrip" or "States01") is totally irrelevant

$>$ here. On the other hand, it would be more interesting to convey
what these programs are actually doing: how

$>$ are the edges added? randomly?, how are they stripped to
critical sets?, etc.

\smallskip
Again and again. The MMPH language and its algorithms and programs
is a {\em sine qua non\/} for generating the MMPHs. How? Randomly?
On p.~2 the referee can read: ``{\bf M2} consists in an automated
random addition of hyperedges\dots'' How are they stripped to
critical sets? Also on p.~2 (left column, bottom): ``A critical
NBMMPH is an NBMMPH which after removing any of its hyperedges
becomes a BMMPH.'' E.g.~1234,4567,789A,ABCD,DEFG,GHI1,29BI,35CE,68FH,F2JA.~is a KS NBMMPH, but not a critical one. By removing the hyperedge F2JA
it becomes a critical KS NBMMPH (Cabello's 18-9). How do we know
that? Because by removing any of the remaining hyperedges it stops
being an NBMMPH (contextual) and turns into a BMMPH (noncontextual).

\smallskip
$>$ -- Page 2, col. 2, Fig. 1:

$>$ This figure is hard to interpret. The caption does not
adequately explain, e.g., what do the axes represent, or 

$>$ what are the shaded black stripes. The same applies to Fig 3
and Fig 4 later on.

\smallskip
The referee is bold again. Abscissa = $l$ (number of hyperedges);
ordinate = $k$ (number of vertices). Dots represent ($k,l$).
Consecutive dots (same $l$) are shown as strips. Analogous
figures have been presented in more than 20 papers of ours
and all referees or those who cited or commented on our papers
or to whom the papers were presented to conferences have obviously
immediately grasped what the axes represent since no one has ever
posed such a question.

\smallskip
$>$ -- Page 2, col. 2, l. 13-14:

$>$ "When applying M3 we obtain that Bub’s is the only 49-36 NBMMPH
and that there are no smaller KS ones 

$>$ for the considered vector components." Again, this statement is
unnecessarily vague. What is meant by "for the 

$>$ considered vector components"? Which vector components are being
considered which allow the authors to make

$>$ this statement?

\smallskip

Another failed weaponized pedantry. Vector components are shown in
Fig. 1 ($0,\pm1,\dots,2\omega^2$). 

\smallskip

$>$ This goes back, I believe, to the overly vague explanation given of M3.

\smallskip 
Such comments, I'm sure, do not go anywhere. 

\smallskip 
$>$ -- Page 5, col.1, l. 26-30

$>$ "The methods are especially needed in odd dimensional Hilbert
spaces since, in contrast to even dimensional

$>$ ones, we cannot make use of polytopes, Pauli operators,
qubit states, parities, and other approaches specific

$>$ to qubit spaces". Are analogous methods not available
for qudits? Is there a reason why this is the case?

\smallskip
It is so good that this is the last referee's comment. We cannot
make use of them because they are ``specific to qubit spaces''
for whatever reason. We do not want to speculate about reasons.
It is outside the scope of the article. It suffices that there
no such methods in the literature and this is transparent from
the fact that so far there were only a dozen of MMPHs found in
the odd-dimensional spaces. 

\section*{Appendix 2:  {\em Physical Review Letter}
  Editor's 2nd decision, 2nd referee report, and
  my response to it}

  \subsubsection*{\rm Mladen Pavi\v ci\'c}
  
  \subsection*{Editor's 2nd decision}

  Journal:\quad Physical Review Letters; \quad
  Paper:\quad LP17301; \quad Received:\quad 17Feb2022

  16Sep22 \phantom{16Jun22} Editorial decision

  29Jul22 01Sep22 Review request to referee; report received
  
Date: 14 Sep 2022

Re: LP17301
    Automated generation of arbitrarily many Kochen-Specker and other
    contextual sets in odd dimensional Hilbert spaces
    by Mladen Pavi\v{c}i\'{c} and Norman D. Megill

Dear Prof. Pavicic,

This manuscript has been reviewed by our [2nd] referee. A critique
from the report appears below. Based on this we judge that the work
probably warrants publication in some form, but does not meet the
Physical Review Letters criteria of impact, innovation, and interest.
In accordance with our standard practice, this concludes our review
of your manuscript.

The paper, with revision as appropriate, might be suitable for
publication in one of the topical Physical Review journals or Physical
Review Research. The editors of that journal will make the decision on
publication, and may seek further review.

Yours sincerely,

Robert Garisto (he/him/his)
Editor Physical Review Letters

\bigskip

---------------------------
{\bf Report of the 2nd Referee -- LP17301/Pavicic}
---------------------------

\medskip

The work ``Automated generation of arbitrarily many Kochen-Specker and
other contextual sets in odd-dimensional Hilbert spaces'' by Pavicic
and Megill introduces numerical methods to obtain Kochen-Specker (KS)
sets (or in general contextual sets) in odd dimensions. Their methods
yield a huge number of KS sets in dimensions 3, 5, 7, and 9 that are
not known before. KS set is a set of projectors that cannot be
assigned pre-determined values independent of the context of the
observables in which the projector belongs. KS sets are genuine
signatures of nonclassicality and have found applications in quantum
computation and communication. There are, in general, two ways that
contextual sets are associated with quantum information. Contextuality
is shown to be necessary for quantum advantages over classical systems
in certain aspects, that is, an advantage by quantum systems in
certain aspects of information processing over classical systems
implies some particular proofs of contextuality (for example, Ref. [1]
in the manuscript). On the other hand, it has also been pointed out
that contextual sets are sufficient for quantum advantage in some
particular information processing tasks. In other words, uncovering
any KS set immediately implies the existence of some information
processing tasks with quantum advantages (for example, Ref. [4] in the
manuscript). In the latter sense, findings of new KS sets have
implications for quantum information. It is also fundamentally
interesting to know about novel KS sets as they give rise to
compelling nonclassical correlations. And only a few KS sets are
indeed known so far in odd dimensions. So I think the question
addressed here in this work is important.

Before I present my opinion regarding the suitability of this work for
publishing in PRL, I would like to pose a few queries and suggestions.

1. KS MMPH is defined as the NBMMPH with k(i) =n for all i. If an
NBMMPH does not have coordinatization then the graph has no quantum
realization. However, one may get confused with the terminology of
`KS MMPH' since the ``KS set'' is usually referred to a set that
already has a quantum realization. I would recommend adding one
line in the text to clarify this. Are there examples of non-KS
NBMMPH without known coordinatization?

2. Do non-KS NBMMPHs always give rise to proof of contextuality? In
other words, given any non-KS NBMMPH, is it always possible to propose
a noncontextuality inequality that quantum theory violates?

3. I would request the authors to mention some details about the
software or/and tools used to execute the numerics so that it would be
convenient to study this technique in the future.

4. A huge number of critical KS sets are found, and it would be nice
if some sort of comparison is made among those critical sets. For
instance, one can consider any measure of robustness for contextuality
and study which set is more robust, or, one can see which set has the
highest ratio between the chromatic number of the graph and n (the dim
of the Hilbert space). This is just a suggestion.

Overall I find this work interesting. But, there is a substantial
limitation of this work. The proposed numerical methods are
computationally demanding for higher dimensions, and the KS sets are
found only in a few small dimensions. Thus the proposed methods for
automated generation of contextual sets are applicable to a certain
extent. This is the reason I think the paper is suitable for
specialized audiences, and I am unable to give a reason why the
manuscript deserves publication in PRL. Given the queries mentioned
above are appropriately addressed, this work can be published in a
specialized journal.

\bigskip
---------------------
{\bf Author's response to the Report of the 2nd Referee --
  LP17301/Pavicic}
---------------------

\medskip

In the revised version of the article ( available at
https://arxiv.org/abs/2202.08197 ) I made some changes 
suggested by the referee. In my response below, I
indicate such changes by ``*****************''. Referee's
comments I indicate by ``$>$''.

$>$ The work ``Automated generation of arbitrarily many Kochen-Specker
and other contextual sets in odd-

$>$ dimensional Hilbert spaces'' by Pavicic and Megill introduces
numerical methods to obtain Kochen-Specker (KS)

$>$ sets (or in general contextual sets) in odd dimensions. Their
methods yield a huge number of KS sets in

$>$ dimensions 3, 5, 7, and 9 that are not known before. KS set is a
set of projectors that cannot be assigned

$>$ pre-determined values independent of the context of the
observables in which the projector belongs. KS sets

$>$ are genuine signatures of nonclassicality and have found
applications in quantum computation and

$>$ communication. There are, in general, two ways that contextual
sets are associated with quantum information.

$>$ Contextuality is shown to be necessary for quantum advantages
over classical systems in certain aspects, that is,

$>$ an advantage by quantum systems in certain aspects of information
processing over classical systems implies

$>$ some particular proofs of contextuality (for example, Ref. [1]
in the manuscript). On the other hand, it has also

$>$ been pointed out that contextual sets are sufficient for quantum
advantage in some particular information

$>$ processing tasks. In other words, uncovering
 any KS set immediately implies the existence of some information

$>$ processing tasks with quantum advantages (for example, Ref. [4]
 in the manuscript). In the latter sense, findings

$>$ of new KS sets have implications for quantum information. It is
also fundamentally interesting to know about

$>$ novel KS sets as they give rise to compelling nonclassical
correlations. And only a few KS sets are indeed known

$>$ so far in odd dimensions. So I think the question addressed here
in this work is important.

\bigskip

In contrast to the 1st referee, the present referee is
at home with both the hypergraph theory and contextual sets.
Her/his understanding and evaluation of the results achieved in
the paper are accurate---up to the last point below---which
is not. 

\bigskip 
$>$ Before I present my opinion regarding the suitability of this
work for publishing in PRL, I would like to pose a

$>$ few queries and suggestions.

\smallskip 
$>$ 1. KS MMPH is defined as the NBMMPH with $k(i)=n$ for all
$i$. If an NBMMPH does not have coordinatization

$>$ then the graph has no quantum realization. However, one may get
confused with the terminology of `KS MMPH' 

$>$ since the ``KS set'' is usually referred to a set that already
has a quantum realization. I would recommend adding

$>$ one line in the text to clarify this. Are there examples of
non-KS NBMMPH without known coordinatization?

\bigskip
**** In the 2nd paragraph of the left column on p. 2, the following
sentences are added: ****

\smallskip
************** The assignments of 0s and 1s do not require
a coordinatization ... **************

\smallskip
************** An example of a non-KS NBMMPH without known
coordinatization is the 33-27 in Fig.~1(d) in the Supplemental
Material (SM). **************

\bigskip
$>$  2. Do non-KS NBMMPHs always give rise to proof of
contextuality? In other words, given any non-KS

$>$ NBMMPH, is it always possible to propose a noncontextuality
inequality that quantum theory violates?

\bigskip
**** Yes. See Table IX in [14]. ****

\bigskip

$>$  3. I would request the authors to mention some details
about the software or/and tools used to execute the

$>$ numerics so that it would be convenient to study this technique
in the future.

\bigskip
**** What is of the primary interest here are the algorithms. They
are universal and they hold in any dimension. I added the
following paragraph to the left column on p. 1: ****

\medskip
************** In this paper, we offer universal and general
algorithms for automated generation of arbitrarily many contextual
sets in any dimension. In contrast to them, the programs we wrote
to implement them are computationally demanding and therefore, here,
we use them to generate sets that have not been generated so far:
billions of KS and contextual non-KS sets in $n=3,5,7,9$-dim spaces.
The programs themselves are freely available from our repository and
technical details of their previous versions are given in [5-9].
**************

\bigskip

$>$ 4. A huge number of critical KS sets are found, and it would be
nice if some sort of comparison is made among

$>$ those critical sets. For instance, one can consider any measure
of robustness for contextuality and study which

$>$ set is more robust, or, one can see which set has the
highest ratio between the chromatic number of the graph

$>$ and $n$ (the dim of the Hilbert space). This is just a suggestion.

\bigskip

**** I have done this in [14]. **** 

\bigskip

$>$  Overall I find this work interesting. But, there is a
substantial limitation of this work. The proposed numerical

$>$ methods are computationally demanding for higher dimensions,
and the KS sets are found only in a few small

$>$ dimensions. Thus the proposed methods for automated generation
of contextual sets are applicable to a certain

$>$  extent. This is the reason I think the paper is suitable for
specialized audiences, and I am unable to give a reason

$>$ why the manuscript deserves publication in PRL. Given the
queries mentioned above are appropriately addressed,

$>$ this work can be published in a specialized journal.

\bigskip
 
**** The referee misunderstood the main accomplishment of the paper.
Our algorithm and method are applicable to any dimension and are
therefore universal and general---they are not ``computationally
demanding''---our programs based on them are. The present
computational barrier which prevents us from providing outputs in
higher dimensions is not their ``limitation'' in any way that might
diminish their importance for the PRL readership. Following our
algorithms one can obtain outputs in higher dimensions by making
use of more powerful supercomputers or new linear programs one
might invent in the future. Also, a quantum computer would allow
for outputs in any dimension. Is the discovery of the DNA structure
unimportant just because we still cannot calculate how to
manipulate all of its sequences? ****

\bigskip

**** The algorithm we propose is the only universal method for
generating arbitrary contextual sets known in the literature
and that, in my opinion, warrants its publication in PRL.
I hope the referee will concur.****

\section*{Appendix 3:  {\em Physical Review Letter}
  Editor's 3rd decision}

  \subsubsection*{\rm Mladen Pavi\v ci\'c}
  
  \subsection*{Acknowledgement LP17301 Pavicic}
  
  Date: Mon, 19 Sep 2022 11:49:44 +0000

  From: prltex@aps.org

Subject: Acknowledgement LP17301 Pavicic

Re: LP17301
    Automated generation of arbitrarily many Kochen-Specker and other
    contextual sets in odd dimensional Hilbert spaces
    by Mladen Pavi\v{c}i\'{c} and Norman D. Megill

Dear Prof. Pavicic,

We have successfully generated output of the above manuscript that you
sent for resubmission.  The manuscript has been forwarded for further
processing.

Yours sincerely,

Editorial Systems
Physical Review Letters

  \subsection*{Editor's 3rd decision}

Date: Fri, 30 Sep 2022 14:53:08 +0000

Subject: Your manuscript LP17301 Pavicic

Re: LP17301

Dear Prof. Pavicic,

This is in response to your resubmission.

Reopening review at this stage would be an extraordinary step, and we
see insufficient cause for that.

Thus, we turn the matter back to you. We suggest that you submit to a
more specialized journal, perhaps Physical Review A.

Yours sincerely,

Robert Garisto (he/him/his)
Editor
Physical Review Letters

\section*{Appendix 4: Comment  on {\em Physical
    Review Letter\/} Editor's 3rd decision}

  \subsubsection*{\rm Mladen Pavi\v ci\'c}

Date: Sat, 1 Oct 2022 23:25:42 +0200

Subject: To the exclusive attention of Dr. Garisto - LP17301 Pavicic

\bigskip
Dear Dr. Garisto,

\medskip
On Fri, Sep 30, 2022 you wrote:

\smallskip
$>$ Reopening review at this stage would be an extraordinary
step, and we see insufficient cause for that.

\bigskip

What is extraordinary here is that I have received no comments
on my responses and changes from either of the referees.

\bigskip

With the best regards, 

\smallskip
Mladen Pavicic.

\section*{Appendix 5:  {\em Quantum} Journal Editor's
  decision and my response to it}

\subsection*{Publication Decision from Quantum}

\subsubsection*{\em Automated Generation of Arbitrarily Many
  Kochen-Specker and Other Contextual Sets in Odd
  Dimensional Hilbert Spaces}

Decision made on October 20th, 2022
Editorial board's determination: Reject

\subsubsection*{Comments from the Admin}

Dear Mladen Pavicic,

Thank you for submitting to Quantum. Unfortunately, we have
to inform you that the editorial board of Quantum have decided
not to publish your paper “Automated Generation of Arbitrarily
Many Kochen-Specker and Other Contextual Sets in Odd Dimensional
Hilbert Spaces”. This decision is due to the following reason:

We have noticed that you have made public, in arXiv:2202.08197,
the referee reports which you received for this manuscript.
The reports were anonymous and formed a confidential communication
between author and editor. By making them public, including your
denigrating comments on the scientific expertise of one of the
referees, you have breached this confidentiality. Moreover, by this
conduct you have put an undue pressure on the refereeing process at
Quantum, since potential referees can expect a similar action in
response to their own reports, if these happen to be critical as
well. (This case of “pressure” by the authors on editors or referees
is explicitly mentioned in our editorial policies, as a ground for
desk rejection.)

For this reason we are not able to consider you paper for
publication in Quantum.

We hope that this quick response will allow you to submit to another
journal, and we look forward to receiving your future work.
Thank you for supporting our efforts towards open, community-led and
fair science publishing.

Best regards,

Lukas Schalleck (Admin)
on behalf of the Editorial Board of Quantum

This rejection letter is final and this discussion is now closed.
Under certain conditions authors can appeal against editorial decisions.

\subsection*{My response to the Decision from Quantum}

Mladen Pavicic Oct 25, 2022 - 12:12 pm CEST

\bigskip
Dear Mr. Schalleck,

\medskip
Lukas Schalleck (Admin) replied to a discussion on the manuscript:

{\em Automated Generation of Arbitrarily Many Kochen-Specker and
Other Contextual Sets in Odd Dimensional Hilbert Spaces. 
(Quantum)}

$>$ Further to this decision and as previously indicated, we would
like to inform you that desk-rejections are not

$>$ eligible for appeals.

OK. But you write: ``This case of ``pressure'' by the authors on
editors or referees is explicitly mentioned in our editorial
policies, as a ground for desk rejection.''

Where exactly? I was unable to spot it in your documents. Please,
help me with that and direct me, as an author, to the relevant
link and lines in it.

With the best regards,

M. Pavicic.

\subsection*{Admin's response}

Lukas Schalleck (admin) Oct 25, 2022 - 12:24 pm CEST

\bigskip
Dear Mladen Pavicic,

\medskip
Thanks for reaching out again. Happy to help, of course: this may be
found under ``Acceptance criteria'' $\to$ ``Editorial pre-selection''
on the following page:
{\tt https://quantum-journal.org/editorial-policies/}

The pertinent lines state: ``External pressure by the authors on
editors or referees outside the due editorial process may also
result in editorial rejection.''

Best regards,
Lukas Schalleck

\subsection*{My response to Admin's response}

Mladen Pavicic, Oct 25, 2022 - 1:26 pm CEST

\bigskip
Dear Mr. Schalleck,

\medskip
I fail to grasp how my publication of {\bf anonymous} previous
reports can be possibly rendered as: ``External pressure by the
authors on editors or referees'' when there is a standard procedure
to send previous reports to next referees of a paper in most
journals and to send previous reports to other journals of the same
publisher.

By these standards, insight into previous reports can only
be helpful and by no means a ``pressure.''

With the best regards,

M. Pavicic.

\section*{Appendix 6:  {\em Physical Review A}
  Editor's letter, 3 referee reports, 
  my responses to them, and Editor's decision}

{\em Physical Review A} has taken the manuscript and the reports
over from the {\em Physical Review Letters} 

\subsection*{Editor's letter}

Date: Tue, 13 Dec 2022

Re: LP17301AR

Automated generation of arbitrarily many Kochen-Specker and other
 contextual sets in odd-dimensional Hilbert spaces
    by Mladen Pavi\v{c}i\'{c} and Norman D. Megill

Dear Prof. Pavicic,

The above manuscript has been reviewed by three of our referees.
Comments from the reports appear below for your consideration.

When you resubmit your manuscript, please include a summary of the
changes made and a brief response to all recommendations and
criticisms.

Yours sincerely,

Dr. Erin Knutson
Associate Editor
Physical Review A

P.S. We are taking referee A's comments seriously. It will be easy for
the review process to become (more) argumentative at this point, and
we want to avoid that outcome. Please be sure to thoroughly and
politely address their comments, which are designed to improve the
impact and readability of your manuscript.

\subsection*{Three reports}

---------------------------------------------------------------------

Second Report of Referee A -- LP17301AR/Pavicic

---------------------------------------------------------------------

\bigskip
{\bf Overview}

\medskip
This is a resubmission of a manuscript originally submitted to PRL.
In my original referee report I raised some concerns about the paper
and made a number of suggestions for improvement, encouraging the
author to submit a carefully revised version to e.g. PRA. When I
received a revised submission I was convinced this would be an easy
case of recommending acceptance to PRA, with at most only some minor
remarks.

However, I am sad to say that I maintain most of my reservations
about the manuscript. Unfortunately the author seems to have
followed only the last part of the recommendation and resubmitted
an almost identical version to PRA. I was disappointed to see that
they made very little effort to address or engage with almost all of
the points raised in the report.

Even regarding fairly minimal comments (such as a request to clarify
the caption of a figure, see point 11 below), the author simply did
not bother to make the very small, harmless change suggested to
improve readability (instead suggesting in their reply that the
referee should have ``immediately grasped what the axes represent'').
This was of course a very minor and unimportant point in the scheme
of the paper, but I think this reply is indicative of the author’s
failure to take the peer-review process seriously enough and to make
an effort to engage with criticism and suggestions.

More worryingly, their reply to some of the more substantial points
seems to betray some basic misunderstandings. The most shocking is
the insistence on presenting unimportant implementation details,
such as keeping the list of ASCII characters used to encode
hypergraphs in their programs as ``an integral part of the MMPH
language,'' and even doubling down on this. This seems to show a
failure to grasp the distinction between (high-level) algorithms
(which are relevant to the reader) and (low-level) implementation
choices (which are incidental and completely irrelevant). I had
commented that choosing to present such implementation details
obfuscates the actual contributions of the paper---however, this
reply made me think that the distinction is not even clear to them.

As another example, similar remarks apply to the suggestion
regarding the definition of MMPH hypergraphs. The author’s reply
mentions a standard textbook to counter the suggestion raised in
the report, claiming that in that book the definition appears in a
similar form as in their manuscript. Looking at the book, one
actually finds out that this is not the case: in the book the
presentation follows the suggestion I had given in my report. Once
again, no change was made to the manuscript regarding this point.

In the remainder of this review, I go into more detail into the
issues mentioned above and consider the author’s reply point by
point. But to summarize: in addition to the reservations I had
expressed about the adequacy of the way their methods are presented
(which were not addressed), the replies quoted above and the
underlying misconceptions made me have bigger doubts about the
quality of this paper.

For the reasons just outlined, I believe that the manuscript should
not be accepted in its current form. As I still believe that there
is worthwhile work behind this paper, I recommend giving the author
another opportunity to revise the manuscript and resubmitting it as
a PRA Letter, in the hope that they’ll engage with criticism and
suggestions in a more constructive fashion to improve the presentation.

\bigskip
Detailed comments

\medskip
Quotes in pink are from the previous referee report and in blue from
the author’s response.

\begin{enumerate}
\item $<$ {\color{magenta}\em ‘[\dots] It claims that ``development of
    quantum computation and communication, recently shown to be
    supported by contextuality, arguably asks for an abundant supply
    of contextual sets,'' but the authors do not bother to explain
    why such sets might be necessary [\dots]’}

  $>$ {\color{blue}\em ‘On the contrary, the motivation is expressed
    in the second paragraph of the paper: “none of the methods
    [previously employed in the even-dimensional spaces] is available
    in the odd-dimensional spaces. In this paper, we offer several
    methods for automated generation of arbitrarily many KS sets as
    well as contextual non-KS ones in $n=3,5,7,9$ dimensional spaces.”
    The sentence: ``quantum computation and communication \dots\
    arguably asks for an abundant supply of contextual sets'' is
    taken from the abstract which is limited in size. In the very body
    of the article, it is substantiated by the motivation sentences
    cited above.’}

  These sentences motivate why one would look at odd-dimensional
  subspaces given that prior work has focused on even-dimensional
  subspaces. These sentences do {\bf not} address the following
  point I was referring to: why does the ‘development of quantum
  computation and communication’ ask for a requisite/abundant supply
  of contextual sets? I understand that quantum communication and
  computation are ‘supported by contextuality’, but what is the
  actual use of having a large supply of contextual sets? Again,
  I am not claiming that finding such contextual sets is not
  necessary. I am simply stating that the article does not argue
  for this need, beyond a this flimsy reference to contextuality
  supporting communication and communication. More motivation on
  this point would make the manuscript stronger and more broadly
  appealing. Having said that, I think this is a less important
  point than previously for PRL.
\item $<$ {\color{magenta}\em ‘The presentation falls short of the
  standard expected in a research article, especially in a journal
  as good as PRL, in terms of clarity and rigor. Various passages
  are written in a confusing fashion.’}

$>$ {\color{blue}\em ‘We published in Phys. Rev. Lett., Scientific
  Reports, Phys. Rev. A \&\ D, Optics Express, J. Math. Phys.,
  J. Phys. A, J. Opt. Soc. Am. B, Entropy, Phys. Lett. A,
  Found. Phys., Ann. H. Poincare \dots\ over almost half a century,
  and we do not think we wrote this paper in a “confusing fashion.”’}

The authors have produced great research and published excellent
papers over their careers. But as far as I am concerned their prior
publishing record is irrelevant in this context where we’re only
being asked to judge the one manuscript at hand. I fail to
understand the relevance of bringing this up. As for the substance
of the criticism, the overview paragraph (as well as the detailed
comments in my first report) highlight various passages where the
presentation was, in my opinion, confusing. The author would do
better to respond to these concerns instead of
bragging about their publication list.
\item $<$ {\color{magenta}\em ‘For example, it is often unclear from
    the text that a passage is to be taken as the definition of a
    new concept (e.g. when the notion of coordinatization is
    introduced).’}

  $>$ {\color{blue}\em ‘Wiktionary has a single item for:
    coordinatization (plural coordinatizations): (mathematics) ``The
    application of coordinates to a system,'' and in the 3rd paragraph
    of the right column on p.~1 we additionally clarify: ``The MMPHs
    above are defined without a coordinatization, i,e., neither their
    vertices nor their hyperedges are related to either vectors or
    operators.'' So, no ``new concept'' is introduced. The
    ``coordinatization'' is a word from a dictionary the referee
    should have consulted.’}

  The dictionary definition is characteristically vague, as suits a
  dictionary of a natural language. It does not explain what the word
  means in specific contexts where is is used with a precise, formal,
  mathematical meaning. There are many ways one could instantiate
  this broad natural-language definition (‘applying coordinates to a
  system’) in the concrete setting of hypergraphs. ‘Coordinatization
  of a MMPH’ is a concept that is first (i.e. newly) introduced in
  this text in the last paragraph of page 1. There, its meaning is
  not precisely set out. One infers from the way the sentence is
  written (and from one’s knowledge of the English language) that a
  coordinatization of an MMPH is a map from vertices or from
  hyperedges (which?) to vectors or operators (which?!). This is
  sloppy and imprecise. Which is it? Do you mean that a
  coordinatization associates to each vertex a vector or an operator?
  Or are such vectors (or operators) associated to hyperedges instead
  of vertices? Are such associations supposed to satisfy some
  constraints, such as the vectors needing to be orthogonal when they
  belong to the same hyperedge (as one seems to infer in between the
  lines from reading a few sentences ahead)?

  Concepts should be defined precisely and fully in scientific
  (especially mathematical) text. The reader is not supposed to infer
  a definition by trying to reconstruct what was in the authors’
  head from pieces of sentences here and there.
\item $<$ {\color{magenta}\em ‘On the other hand, e.g. in the
  definition of MMPH hypergraph, some extra commentary is written as
  a part of the definition when it does not belong there, namely
  item (v).’}

$>$ {\color{blue}\em ‘In the theory of hypergraphs, e.g., Berge,
  Hypergraphs, North Holland (1989), Ch.~1, Sec.~1, p.~1, 4th
  paragraph it is assumed that ``A hypergraph may be drawn as a set
  of points representing the vertices [and a hyper]edge is
  represented \dots\ by a single closed curve enclosing the vertices
  is a part of the definition.'' There is no particular kind of a
  ``definition'' of ours which would differ from Berge’s or any
  other standardly accepted definition except that we introduce, in
  point (v), a curve passing through vertices instead of Berge’s
  curve which encloses them.’}

$<$ {\color{magenta}\em ‘ ``(v) graphically, vertices are represented
  as dots and hyperedges as (curved) lines passing through them.''
  This is given as one of the conditions in the definition of MMPH
  graphs, but in reality it is not a condition on hypergraphs. It is
  just an explanation of the convention used to draw hypergraphs to
  ease the readability of Figure 2, for example. It should be
  separated from the definition of MMPH hypergraph.’}

$>$ {\color{blue}\em ‘Nowhere do we speak about a condition. It is
  just the point within a definition in the same way in which similar
  points are used in many hypergraph textbooks. We already clarified
  that point (v) above when we compared it with a similar point used
  by Berge, Ch.~1, Sec.~1, p.~1, 4th paragraph.’}

I located a copy of the book mentioned. It is surprising that the
author chose to point to it. The beginning of Chapter 1, which
introduces the notion of hypergraph, is indeed clearly written.
But it does not do what I criticised in this manuscript; instead,
the defining conditions and additional commentary on notational or
pictorial conventions are clearly delineated from each other.
I suggest that the author actually take this passage of the book as
a model.

Berge gives the definition of a hypergraph (‘A {\em hypergraph} on
$X$ is a family \dots\ such that (1) \dots\ (2) \dots\ A simple
hypergraph \dots\ is a hypergraph \dots\ such that (3) \dots’).
Then in the following paragraph Berge goes on to explaining that
elements of $X$ are called vertices etc. as well as setting out
some other conventions. Then the paragraph after that contains the
sentence quoted, explaining the book’s convention for
depicting/drawing hypergraphs. This is most definitely not part
of the definition, which Berge clearly demarcates.

Quite differently, the manuscript reads: ‘An MMPH is a connected
$n$-dim hypergraph $k$-$l$ with $k$ vertices and $l$ hyperedges in
which (i) \dots; (ii) \dots; (iii) \dots; (iv) \dots;
(v) graphically, vertices are represented \dots’. Here, these are
presented as five conditions (on hypergraphs) defining what it
means to be an MMPH hypergraph. Item (v) is not (as in Berge’s book)
presented as what it is: a convention on how to depict MMPH
hypergraphs, rather than an item that is part of the definition,
i.e. of what it means to be an MMPH hypergraph. It should, as in
Berge’s text, be clearly separated.

The author claims that this is never spoken about as a ‘condition’.
But in fact the paragraph defining MMPH hypergraphs is phrased in
a way that places (v) on par with the other conditions defining what
it means to be a hypergraph. And moreover, later in the text one
reads: ‘An MMPH is a special kind of a general hypergraph in the
sense that none of the aforementioned points (i-v) holds for it’,
which implicitly says that (i-v) are to be taken as conditions on
hypergraphs that define what it means to be an MMPH hypergraph.
\item $<$ {\color{magenta}\em ‘The choice of material included in
  the text is odd, and it is inadequate for a journal article.
  At points, the authors choose to give too much information on
  irrelevant implementation details (e.g., the internal ASCII
  encoding of the graphs being manipulated, or the name of
  functions in their computer code).’}

$>$ {\color{blue}\em‘ Again, Berge, Ch.~1, Sec.~1, p.~1, 5th
  paragraph: ``One may also define a hypergraph by its incidence
  matrix, with columns representing the [hyper]edges and rows
  representing the vertices (cf. Figure 1, p. 2).++ We, instead
  of an incidence matrix, make use of MMPH ASCII strings. Some
  kind of codification is unavoidable and our MMPH string
  codification is an integral part of the MMPH language and these
  strings are generated by our algorithms and programs and they
  then serve us as further direct inputs to our algorithms and
  programs whose outputs are all MMPH properties and features
  including graphical representations of the obtained MMPHs. The
  difference between incidence matrices and MMPHs is obvious;
  for instance, the MMPH string of Fig. 2(c) can be written in a
  single line, while the incidence matrix takes 37 columns and 51
  lines. Also an MMPH string can be put in any of our programs
  while no such programs for arbitrary incidence matrices are
  known to us.’}

$<$ {\color{magenta}\em ‘This is the kind of implementation detail
  that is completely irrelevant to the potential readers of this
  paper. Its inclusion is a waste of space that could be used to
  convey much more interesting information about the methods
  implemented. This does not concern any substantial/significant
  algorithmic choice, but simply a matter of convention chosen for
  the internal representation. Its the kind of detail that is
  typically hidden from the user even in the documentation of a
  well-written library, and its even less relevant in the context
  of a journal article, which ought to focus on explaining the
  methods instead.’}

$>$ {\color{blue}\em‘There are no methods which could generate
  the contextual sets in odd dimensional spaces outside the MMPH
  language based on ASCII codification either in our minds or
  anywhere in the literature. The MMPH language is the method.
  A ``method instead'' does not exist.’}

$<$ {\color{magenta}\em‘This is problematic for two reasons.
  First, it detracts from the interest that the manuscript may
  have to a journals audience: most readers would read the paper
  to learn the main ideas underlying the methods so they could
  replicate and adapt them (anyone keen on learning implementation
  details would refer to the codes documentation and/or to the
  code itself )\dots’}

$>$ {\color{blue}\em‘The 2nd sentence seems to contradict itself:
  ``most readers would read the paper to learn the main ideas
  underlying the methods so they could replicate and adapt them,''
  but they would not like ``to learn the code itself.'' Note that
  without the MMPH language the results obtained in the article
  are inaccessible as they were in the last half of century.’}

These replies contributed to seriously aggravate my misgivings,
in particular the claims that the string codification chosen is
‘an integral part of the MMPH language’, or that ‘there are
no methods \dots\ outside the MMPH language based on ASCII
codification’ and that this ‘MMPH language is the method,' or the
outrageous suggestion that ‘without the MMPH language the
results obtained in the article are inaccessible’. These seems to
reveal some very basic misunderstanding on the author’s part of
the distinction between the level of ideas, methods, algorithms
and that of concrete implementation details. The higher level
abstracts from such irrelevant details and it should be conveyed
in a way that is oblivious to it.

For illustration, let me first remark the obvious point that the
use of ASCII characters is totally immaterial. If instead the
author had used Unicode characters or 16-bit integers to encode
hypergraphs, nothing of relevance would change whatsoever (except
that it would take longer for the available ‘characters’ to be
exhausted).

Similarly, the point about a string being written ‘in a single
line while the incidence matrix takes 37 columns and 51 lines’
is misguided. For example, many graph algorithms are typically
presented in terms of incidence matrices (or even more abstractly),
but of course ultimately everything is represented in the computer
as a sequence of bits, which can in theory be displayed in a single
line. One doesn’t usually do this for a reason: such encodings are
cumbersome and distract from the essence of the algorithms. The
great power of abstraction is precisely that it allows one to focus
on the relevant structure at each level, without concerning oneself
with irrelevant, low-level details.

Finally, regarding comment that ‘an MMPH string can be put in any
of our programs while no such programs for arbitrary incidence
matrices are known to us’ let me just reiterate that a research
article is not a manual documenting the usage of a computer
program. In such a context, it would be natural to describe the
format in which the program expects to receive its input, so that
a potential user could interface with the program. But that’s
completely beside the point for a journal article.
\item $<$ {\color{magenta}\em‘I would therefore encourage the
  authors to submit a carefully *revised* version, with a more
  detailed presentation of the main novel technical contributions,
  to a journal such as PRA.’}

$>$ {\color{blue}\em‘Our algorithms do require sophisticated
  programs but that does not make the contributions ``technical,''
  because there is no other way to obtain them.’}

The last sentence makes very little sense. If the authors don’t
recognise any new technical contributions, why are they submitting
a research article?! What does being a technical contribution have
to do with there not being any other way to obtain them? Anyway,
I did not use the word ``technical'' as a criticism in any way.
My point was that it would be good to see more about the actual
new contributions enabling those “sophisticated programs”.
\item $<$ {\color{magenta}\em ‘ ``Development of quantum computation
  and communication, recently shown to be supported by contextuality,
  *arguably asks for an abundant supply of contextual sets*”.
  Contextuality has been shown to underlie quantum advantage in
  various computation and communication tasks and protocols. However,
  the second part of the sentence does not follow directly from this.
  That is, why does the development of quantum computation and
  communication ask for an *abundant supply of contextual sets*? This
  may well be the case, but (1) it is not an obvious consequence of
  the results mentioned from the literature, and (2) the authors fail
  to argue for this. Given that this is presented as the main
  motivation for this work and for its broad appeal, I would expect
  to see a better case made for this.’}

$>$ {\color{blue}\em ‘Since ``what gives quantum computers that
  extra oomph over their classical digital counterparts \dots\ is
  an intrinsic, measurable aspect of quantum mechanics called
  contextuality'' [1,2], we should have more than a few sets per
  dimension. Those recent results are well-known and we considered
  that it is outside the scope of the present paper to elaborate
  on it.’}

I agree that those recent results about contextuality and quantum
advantage are well known and there is no need to elaborate on them.
But it is not clear from any of these results that finding
contextual sets is required or useful.

\item $<$ {\color{magenta}\em ‘ ``and *several* in 5- to 11-dim
    spaces'' This is a rather vague statement: what is meant by
    ``several''?’}
  
$>$ {\color{blue}\em ‘By Webster Dictionary ``several'' means
  ``more than two but fewer than many.'' ’}

This is yet another flippant reply, not to say rude. The dictionary
meaning of “several” is familiar---that was of course not the point
of my comment. Note that this dictionary definition is again
characteristically and unavoidably vague. The boundary between
``several'' and ``many'' is context dependent. Therefore saying
that ``several'' KS sets were found in 5- to 11-dim spaces
conveys very little information: what order of magnitude are the
authors talking about? Is a dozen ``several'' or ``many''?
A hundred? A thousand?
\item $<$ {\color{magenta}\em ‘ ``An MMPH is a special kind of a
  general hypergraph in the sense that none of the aforementioned
  points (i-v) holds for it.'' The meaning of this sentence is
  unclear. What is the referent of the pronoun ``it'' at the end
  of the sentence?’}

$>$ {\color{blue}\em ‘ ``it'' = a general hypergraph. It is clear
  because points (i-v) obviously hold for MMPH.’}

Well, yes, one can infer what the author meant to write. But
grammatically this sentence is far from being clear. It really is
just poor English, but it is astonishing that the author hasn’t
even attempted to rephrase the sentence after this was pointed out.
\item $<$ {\color{magenta}\em ‘ ``Orthogonality between vertices
  in an MMPH space just means that the vertices are contained in
  their hyperedges” This is yet another sentence whose meaning
  isn’t clear. What is meant by “the vertices are contained in
  *their* hyperedges”?’}

$>$ {\color{blue}\em ‘Meaning is clear. Let me put it this way:
  Let ABC be a hyperedge; for vertices A,B,C, the hyperedge ABC
  is *their* hyperedge; so, the vertices A,B,C are mutually
  orthogonal just by being contained in *their* hyperedge. I'm
  convinced the reader will have no problem with the sentence.’}

The same vertex might belong to more than one hyperedge, so it
makes no sense to speak of ‘its’ hyperedge. A vertex is always
contained, by definition, in any of its hyperedges, so the
sentence as it stands is trivial. What the author wanted to say
is that two vertices being orthogonal means that they are both
contained in a common hyperedge. But this is not what the text
says.
\item $<$ {\color{magenta}\em ‘Page 2, col.~2, Fig.~1: This
  figure is hard to interpret. The caption does not adequately
  explain, e.g., what do the axes represent, or what are the
  shaded black stripes. The same applies to Fig 3 and Fig 4
  later on.’}

$>$ {\color{blue}\em ‘Abscissa = $l$ (number of hyperedges);
  ordinate = $k$ (number of vertices). Dots represent $(k,l)$.
  Consecutive dots (same $l$) are shown as strips. Analogous
  figures have been presented in more than 20 papers of ours
  and all the referees and all from many conferences have
  immediately grasped what the axes represent.’}

I was disappointed to see no change made to the caption to
clarify this point, which could have been very easily addressed.
\end{enumerate}

\bigskip
---------------------------------------------------------------------

Second Report of Referee B -- LP17301AR/Pavicic

---------------------------------------------------------------------

\bigskip
The authors have addressed my comments in the revised version of the
manuscript. A couple of comments are addressed in a later work,
Ref.[14]. I think the current version is suitable for publication in
PRA as a Letter.

The phrase 'do not require' is repeated in the added sentence (2nd
paragraph on page 2).

\bigskip
---------------------------------------------------------------------

Report of the Third Referee -- LP17301AR/Pavicic

---------------------------------------------------------------------

\bigskip
Having read the manuscript and the previous referees' report, I feel
the authors have successfully attended all the concerns of the
referees and this manuscript, containing much new and important
material, should be published in PRA.

The authors state that these results may have future applications in
quantum computing or communication. These hopes are often mentioned in
quantum mechanics manuscript, but no concrete idea is presented. Some
further discussion would be useful.

On page 2 second paragraph line 14 the phrase ``do not require'' is
repeated.

\subsection*{My responses to the Editor and to the
  three reports}

Date: Tue, 13 Dec 2022

\smallskip
Response to Editor's comments on LP17301AR

\medskip
Dear Dr. Knutson,

\dots

In my response to referee A's report I'm now as polite and
cooperative as humanly possible.

\dots

With the best regards,

Mladen Pavicic

\bigskip 
---------------------------------------------------------------------

Response to the Second Report of Referee A -- LP17301AR/Pavicic

---------------------------------------------------------------------

\bigskip
I thank the referee for her/his benevolent attempts to convince
me to ``engage with [her/his] criticism and suggestions in a more
constructive fashion to improve the presentation.'' The referee
made use of the first page of her/his reports to elaborate on
her/his conviction and since real changes are at stake here,
I'm dwelling on them straight away. Here the opinion of the third
referee ``Having read the manuscript and the previous referees'
report, I feel the authors have successfully attended all the
concerns of the referees and this manuscript, containing much
new and important material, should be published in PRA''
suggests that I would hopefully be able to successfully meet
all suggestions put forward by the present referee. 
Note that the changes suggested on p.~1 of the reports are
specified in its Detailed comments to which I respond below. 

\medskip
Pink quotes from the previous referee report are prefixed by $>>>$,
from my previous blue answers by $>>$, and from the present report
by $>$ (I'm writing my response in the plain text mode.)

\medskip
$>$ 1. why does the ‘development of quantum computation and
communication’ ask for a requisite/abundant supply
$>$ of contextual sets? I understand that quantum communication
and computation are ‘supported by contextuality’,
$>$ but what is the actual use of having a large supply of
contextual sets? Again, I am not claiming that finding such\break
$>$ contextual sets is not necessary. I am simply stating
that the article does not argue for this need, beyond this\break
$>$ flimsy reference to contextuality supporting communication
and communication. More motivation on this point would
$>$ make the manuscript stronger and more broadly appealing.
Having said that, I think this is a less important
$>$ point than previously for PRL.

\medskip
The referee is right. I added the following paragraph to
the left column on p.~1: 

``Since the quantum communication and computation are supported by
contextuality [1,2], the actual potential use of a large supply of
contextual sets is twofold. First, quantum computation algorithms
which would rely on contextual sets would arguably rely on a
variety of such sets and on their automated generation. Second,
structural properties of contextual sets differ according to their
coordinatization, parities, dimensions, sizes, etc., and that can
lead us to better understanding and applications of the sets.''

\medskip
$>$ 2. The author would do better to respond to these concerns
instead of bragging about their publication list.

\medskip
I'll do my best.

\medskip
$>$ 3. ‘Coordinatization of a MMPH’ is a concept that is first
(i.e., newly) introduced in this text in the last\break
$>$ paragraph of page 1. There, its meaning is not precisely
set out. One infers from the way the sentence is written\break
$>$ that a coordinatization of an MMPH is a map from vertices
or from hyperedges (which?) to vectors or operators\break
$>$ (which?!). This is sloppy and imprecise. Which is it? Do
you mean that a coordinatization associates to each\break
$>$ vertex a vector or an operator?

\medskip
Yes. 

\medskip
$>$ are such vectors (or operators) associated to hyperedges
instead of vertices? Are such associations supposed to
$>$ satisfy some constraints, such as the vectors needing to be
orthogonal when they belong to the same hyperedge?

\medskip
Yes. I added the following sentence ``The meaning of
coordinatization is that a vector is assigned to each vertex
and that all vectors assigned to vertices belonging to a common
hyperedge are orthogonal to each other.'' as the 2nd sentence
of that paragraph which is now the 2nd paragraph on p. 2.

\medskip
$>$  4. the manuscript reads: ‘An MMPH is a connected n-dim
hypergraph $k$-$l$ with $k$ vertices and $l$ hyperedges in
$>$ which (i) \dots; (ii) \dots; (iii) \dots; (iv) \dots;
(v) graphically, vertices are represented \dots’. Here, these
are presented as
$>$ five conditions (on hypergraphs) defining what it means to
be an MMPH hypergraph. Item (v) is not presented as
$>$ what it is: a convention on how to depict MMPH hypergraphs,
rather than an item that is part of the definition,
$>$ i.e., of what it means to be an MMPH hypergraph. It should be
clearly separated.

\medskip
The referee is right. (v) is not a ``condition'' and now
I clearly separate ``graphically, vertices are represented \dots''
from the conditions ``(i)-(iv)'' and I don't denote it as ``(v).''
I changed the previous last sentence of the 2nd paragraph of
the right column on p.~1 ``\dots; (v) graphically, \dots'' so as
to read : \dots\ Graphically, \dots" 

\medskip
$>$  later in the text one reads: ‘An MMPH is a special kind of a
general hypergraph in the sense that none of the\break
$>$  aforementioned points (i-v) holds for it’, which implicitly
says that (i-v) are to be taken as conditions on hypergraphs
$>$ that define what it means to be an MMPH hypergraph.

\medskip 
The referee is right. I changed "(i-v)" so as to read "(i-iv)." 

\medskip
$>$  5. \dots\ seriously aggravate my misgivings, in particular
the claims that the string codification chosen is ‘an\break
$>$ integral part of the MMPH language’, or that ‘there are no
methods \dots\ outside the MMPH language based on\break
$>$ ASCII codification’ and that this ‘MMPH language is the
method,’ or the outrageous suggestion that ‘without\break
$>$ the MMPH language the results obtained in the
article are inaccessible’.

\medskip
The referee is right. The sentences are too strong. But the
sentences ``an integral part of the MMPH language'' or ``there
are no methods \dots\ outside the MMPH language based on ASCII
codification'' or ``without the MMPH language the results
obtained in the article are inaccessible'' do not appear in the
paper and I completely withdraw such opinions expressed in my
previous response to the previous referee's report. 

\medskip
$>$  These seems to reveal some very basic misunderstanding on the
author’s part of the distinction between the\break
$>$ level of ideas, methods, algorithms and that of concrete
implementation details.

\medskip
Yes, I admit that we ``misunderst[ood] \dots\ the distinction
between the level of ideas, methods, algorithms and that of
concrete implementation details.'' And in particular the referee is
right that ``The higher level abstracts from such irrelevant details
and it should be conveyed in a way that is oblivious to it.''
Therefore I'm ready to, along these lines, adopt the suggestions
made by the referee and implement the following ``concrete
implementation details:''

\medskip
$>$ For illustration, let me first remark the obvious point that
the use of ASCII characters is totally immaterial.\break
$>$ If instead the author had used Unicode characters or 16-bit
integers to encode hypergraphs, nothing of relevance\break
$>$ would change whatsoever.

\medskip
The referee is quite right. We now added the following
paragraph at the top of p.~2:

``Of course, instead of ASCII characters we could have
used Unicode characters or 16-bit integers but 20 years
ago we decided to proceed with the ASCII characters
to encode MMPH strings and design our algorithms and
programs which in turn yield all properties and features
of MMPHs as well as their figures within the MMPH
language. All our papers since 2000 [15] make use of the
ASCII characters for the purpose.''

\medskip
$>$  Similarly, the point about a string being written ``in a
single line while the incidence matrix takes 37 columns\break
$>$ and 51 lines is misguided.''

\medskip
I took out the paragraph completely.

\medskip
$>$ Finally, regarding comment that ``an MMPH string can be put in
any of our programs while no such programs\break
$>$ for arbitrary incidence matrices are known to us'' let me just
reiterate that a research article is not a manual\break
$>$ documenting the usage of a computer program. In such a context,
it would be natural to describe the format in\break
$>$ which the program expects to receive its input, so that a
potential user could interface with the program.\break
$>$ But that’s completely beside the point for a journal article.

\medskip
The referee is quite right. The mentioned sentence is not any
more in the paper and actually neither ``incidence'' nor ``matrix''
nor ``matrices'' do appear in the paper and the ``incidence
matrices'' are not even mentioned in the present revised version
of the paper. 

\medskip
$>$ 6. $>>>$ ``I would therefore encourage the authors to submit a
carefully *revised* version, with a more detailed
$>>>$ presentation of the main novel technical contributions,
to a journal such as PRA.''

$>>$  ``Our algorithms do require sophisticated programs but that does
not make the contributions ``technical,''\break
$>>$ because there is no other way to obtain them.''

$>$ The last sentence makes very little sense. If the authors don’t
recognise any new technical contributions, why\break
$>$ are they submitting a research article?!  What does being a
technical contribution have to do with there not being\break
$>$ any other way to obtain them? Anyway, I did not use the word
``technical'' as a criticism in any way. My point\break
$>$ was that it would be good to see more about the actual
new contributions enabling those “sophisticated programs”.

\medskip
The referee is quite right. I'm now submitting a ``carefully
*revised* version.'' As for ``there is no other way to obtain
them,'' I meant: {\em for us\/}---after more than 20 years of
our usage of our algorithms and programs. Anyhow, the sentence
does not appear in the paper. As for referee's last sentence,
we now point the role of algorithms at several places in the
paper and we also added the following paragraph to p. 2, right
column, 4th paragraph from the top:
``We carry out methods M1-M3 and by means of our programs MMPSstrip
(for stripping and adding hyperedges), States01 (for verifying the
contextuality), MMPShuffle (for reorganizing MMPHs), and ONE (for
evaluating the structural properties of MMPHs [14]).''

\medskip
$>$  7. I agree that those recent results about contextuality and
quantum advantage are well known and there is no
$>$ need to elaborate on them. But it is not clear from any of
these results that finding contextual sets is required or
$>$ useful.

\medskip
The referee is right. I discussed that under point 1 above and
added a relevant paragraph to the left column on p.~1 (also
cited above under point 1). 

\medskip
$>$  8. $>>>$ ``and *several* in 5- to 11-dim spaces''
This is a rather vague statement: what is meant by “several”?\break
$>$ Is a dozen ``several'' or ``many?'' A hundred? A thousand?

\medskip
The referee is right. The sentence now reads:
``\dots, four in the 5-dim space [11],[12, Supp. Material],
five in the 7-dim space [11, 13],[12, Supp. Material],
two in the 9-dim space [12, Supp. Material],
and two in the 11-dim space [12, Supp. Material]."

\medskip
$>$  9. $>>>$ ` ``An MMPH is a special kind of a general hypergraph
in the sense that none of the aforementioned\break
$>>>$ points (i-v) holds for it.'' The meaning of this sentence is
unclear. What is the referent of the pronoun ``it'' at
$>>>$ the end of the sentence?’

$>$  Well, yes, one can infer what the author meant to write. But
grammatically this sentence is far from being clear.

\medskip
The referee is right. The sentence now reads:
``An MMPH is a special kind of a general hypergraph in the sense
that none of the aforementioned points (i-iv) holds for such a
hypergraph.''

\medskip
$>$  10. $>>>$ ‘ ``Orthogonality between vertices in an MMPH space
just means that the vertices are contained in
$>>>$ their hyperedges'' This is yet another sentence whose meaning
isn’t clear. What is meant by ``the vertices are
$>>>$ contained in *their* hyperedges?'' 

$>$  What the author wanted to say is that two vertices being
orthogonal means that they are both contained in\break
$>$ a common hyperedge. But this is not what the text says.

\medskip
The referee is right. The sentence now reads: ``Orthogonality
between vertices in an MMPH space just means that they are
contained in common hyperedges.''

\medskip
$>$ 11. $>>>$  Page 2, col.~2, Fig.~1: This figure is hard
to interpret. The caption does not adequately explain, e.g.,
$>>>$ what do the axes represent, or what are the shaded
black stripes. The same applies to Fig 2 and Fig 3 later on.

$>$ I was disappointed to see no change made to the caption to
clarify this point, which could have been very easily
$>$ addressed.

\medskip 
The referee is right. I now added the following sentences
to the caption of Fig.~1: ``Abscissa is $l$ (number of hyperedges);
ordinate is $k$ (number of vertices). Dots represent $(k,l)$.
Consecutive dots (same $l$) are shown as strips. The same applies
to Figs.~2 and 3.''

\bigskip
-----------------------------------------------------------------

Response to the Second Report of Referee B -- LP17301AR/Pavicic

-----------------------------------------------------------------

\bigskip
$>$ The authors have addressed my comments in the revised version
of the manuscript. A couple of comments are
$>$ addressed in a later work, Ref.~[14]. I think the current
version is suitable for publication in PRA as a Letter. The
$>$ phrase `do not require' is repeated in the added sentence
(2nd paragraph on page 2).

\medskip
I thank the referee. The typo is corrected.

\bigskip
-----------------------------------------------------------------

Response to Report of the Third Referee -- LP17301AR/Pavicic

-----------------------------------------------------------------

\bigskip
$>$  Having read the manuscript and the previous referees' report,
I feel the authors have successfully attended all\break
$>$ the concerns of the referees and this manuscript, containing much new and
important material, should be published
$>$ in PRA.

\medskip
I thank the referee. 

\medskip
$>$  The authors state that these results may have future
applications in quantum computing or communication.\break
$>$ These hopes are often mentioned in quantum mechanics
manuscript[s], but no concrete idea is presented. Some\break
$>$ further discussion would be useful.

\medskip
A discussion is now included in the bottom paragraph of the left
column on p. 1. It reads: ``Since the quantum communication and
computation are supported by contextuality [1, 2], the actual
potential use of a large supply of contextual sets is twofold.
First, quantum computation algorithms which would rely on
contextual sets would arguably rely on a variety of such sets
and on their automated generation. Second, structural
properties of contextual sets differ according to their
coordinatization, parities, dimensions, sizes, etc., and that can
lead us to better understanding and applications of the sets.''

\medskip
$>$  On page 2 second paragraph line 14 the phrase
``do not require'' is repeated.

\medskip
The typo is corrected.

\subsection*{Editor's decision---on Dec 16; my responses
to referee's reports above were submitted on Dec 15}

Acceptance LP17301AR Pavicic

From	pra@aps.org

Date	2022-12-16 20:25

\medskip 
{\em Automated generation of arbitrarily many Kochen-Specker and other
  contextual sets in odd-dimensional Hilbert spaces\/}
by Mladen Pavi\v{c}i\'{c} and Norman D. Megill

{\parindent=0pt
\medskip
Dear Prof. Pavicic,

\medskip
We are pleased to inform you that your manuscript has been accepted
for publication as a Letter in Physical Review A.\break}

Yours sincerely,

\medskip
Davide Girolami, Associate Editor, Physical Review A

\bigskip \bigskip 
{\bf {\em Physical Review A\/} Twitter of the paper:}

https://twitter.com/PhysRevA/status/1611369522254041091

\end{widetext}

\end{document}